\RequirePackage{fix-cm}
\documentclass[smallextended]{svjour3}       
\smartqed  
\usepackage{graphicx}
\usepackage{amssymb}
\usepackage{amsmath}
\usepackage{bbm}
\usepackage[sort&compress, square, numbers]{natbib}
\makeatletter

\begin{document}
\title{
A Non-Local Means Filter for Removing
the Poisson Noise
}


\author{Qiyu JIN         \and
        Ion Grama        \and
        Quansheng Liu
}


\institute{Qiyu JIN \at Institute of Image Processing and Pattern Recognition, Shanghai Jiao Tong University, No. 800 Dongchuan Road, Minhang District, Shanghai 200240, China\\
 UMR 6205, Laboratoire de Math��matiques de Bretagne Atlantique,
              Universit\'{e} de Bretagne-Sud, Campus de Tohaninic, BP 573,
56017 Vannes, France \\
Universit\'{e} Europ\'{e}enne de Bretagne, France\\
Jiangsu Engineering Center of Network Monitoring, Nanjing University of Information Science $\&$ Technology, Nanjing 210044, China\\
              \email{qiyu.jin2008@gmail.com}           
           \and
           Ion Grama \at UMR 6205, Laboratoire de Math��matiques de Bretagne Atlantique,
              Universit\'{e} de Bretagne-Sud, Campus de Tohaninic, BP 573,
56017 Vannes, France\\
Universit\'{e} Europ\'{e}enne de Bretagne, France\\
              \email{ion.grama@univ-ubs.fr}
           \and
           Quansheng Liu \at UMR 6205, Laboratoire de Math��matiques de Bretagne Atlantique,
              Universit\'{e} de Bretagne-Sud, Campus de Tohaninic, BP 573,
56017 Vannes, France\\
Universit\'{e} Europ\'{e}enne de Bretagne, France\\
 School of Mathematics and Computing Sciences, Changsha University of Science and Technology, Changsha
410076, China\\
              \email{quansheng.liu@univ-ubs.fr}
}

\maketitle

\begin{abstract}
A new image denoising algorithm to deal with the Poisson noise model is given, which
is based on the idea of Non-Local Mean. By using the  "Oracle" concept, we  establish a theorem to show that  the Non-Local Means Filter can effectively deal with Poisson noise with some modification. Under the theoretical result, we construct our new algorithm called Non-Local Means Poisson Filter and demonstrate in theory that the filter converges at the usual optimal rate. The filter is as simple as the classic Non-Local Means and the simulation results show that our filter is very competitive.
\keywords{
Non-Local Means\and Mean Square Error\and Poisson noise\and "Oracle" estimator}
\end{abstract}

\section{Introduction}
Noise is inevitable in any image device. A digital imaging system consists
of an optical system followed by a photodetector and associated electrical
filters. The photodetector converts the incident optical intensity to a
detector current, i.e. photons to electrons. During the process, the true
signals are contaminated by many different sources of noise. The Poisson
noise appears in low-light conditions when the number of collected photons
is small, such as night vision, medical imaging, underwater imaging,
microscopic imaging, optical microscopy imaging and astronomy imaging. Such
a noise is signal-dependent, and requires to adapt the usual denoising
approaches.

The key challenge in Poisson intensity estimation problems is that
 the variances of the observed counts are different. As a result, many
methods are introduced to transform the Poisson distributed noise to the
data approximately Gaussian and homoscedastic. These methods are called
Variance Stabilizing Transformations (VST), such as Anscombe root transformation
(1948 \cite{ANSCOMBE1948TRANSFORMATION}, and 1993 %
\cite{borovkov2000estimates}), multiscal VSTs (2008 %
\cite{ZHANG2008WAVELETS}), conditional variance stabilization (CVS) (2006 %
\cite{jansen2006multiscale}), or Haar-Fisz transformation (2004 %
\cite{FRYZLEWICZ2004HAAR} and 2007 \cite{FRYZLEWICZ2007GOES}). Then we can
deal with these data as Gaussian noise. Second, the
noise is removed using a conventional denoising algorithm for additive white
Gaussian noise, see for example
Buades, Coll and Morel (2005 \cite{buades2005review}),
Kervrann (2006 \cite{kervrann2006optimal}), Aharon and Elad and Bruckstein
(2006 \cite{aharon2006rm}), Hammond and Simoncelli (2008 %
\cite{hammond2008image}), Polzehl and Spokoiny (2006 %
\cite{polzehl2006propagation}), Hirakawa and Parks (2006 %
\cite{hirakawa2006image}), Mairal, Sapiro and Elad (2008 %
\cite{mairal2008learning}), Portilla, Strela, Wainwright and
Simoncelli (2003 \cite{portilla2003image}), Roth and Black (2009 %
\cite{roth2009fields}), Katkovnik, Foi, Egiazarian, and Astola (2010 %
\cite{Katkovnik2010local}), Dabov, Foi, Katkovnik and Egiazarian (2006 %
\cite{buades2009note}),
 Abraham, Abraham, Desolneux and  Li-Thiao-Te (2007 \cite{Abraham2007significant}), and Jin, Grama and Liu (2011 \cite{JinGramaLiuowf}).
After denoising, some inverse
transformations, like Exact Unbiased Inverse (EUI) (2009 %
\cite{MAKITALO2009INVERSION} and 2011 \cite{makitalo2011optimal}), are
applied to the denoised signal, obtaining the estimate of the signal of
interest. Many authors restore the Poisson noise by this type of methods
with a three-step procedure (see %
\cite{boulanger2008non,ZHANG2008WAVELETS,lefkimmiatis2009bayesian,luisier2010fast}
).

Maxmum Likelihood (ML) estimation (1996 \cite{van1996comparing}, 2009%
\cite{moon2009three}) and Similarity Measure (SM) (2006 %
\cite{alter2006intensity}) are also found to be effective since they can
account for the special properties of the Poisson distribution. Others
methods like as Complexity-Penalized Likelihood Estimation (CPLE) (2000 %
\cite{nowak2000statistical}, 2005 \cite{kolaczyk2005multiscale}) and
Total Variation (TV) seminorm (2009 \cite{beck2009fast}), have been
introduced to deal with the Poisson noise.
Le et al. (\citep{le2007variational}) have adapted the successful ROF model for total
variation regularization to deal with Poisson noise. The gradient descent iteration
for this model replaces the regularization parameter
with a function.

The Non-Local Means Filter has been proposed by Buades \textit{et al} (2005 %
\cite{buades2005review}) to denoise images damaged by additive white Gaussian noise. It
is based on the similarity phenomenon existing very often in natural images,
and assumes that there is enough redundant information (pixels having
identical noise-free value) in the image to reduce the noise significantly.
This filter is known to efficiently reduce the noise and to preserve
structures. Some authors (see 2008 \cite{boulanger2008non}, 2010 %
\cite{deledalle2010poisson}) combine the Non-Local Means method with other
methods to restore the Poisson noise. Deledalle et al. (2010 \citep{deledalle2010poisson}) proposed an extension of the Non-Local Means   for images damaged by Poisson
noise. It is based on probabilistic similarities to compare
noisy patches and patches of a pre-estimated image.

In this paper, a new image denoising algorithm to deal with the Poisson noise model is given, which
is based on the idea of Non-Local Mean. Our main idea is as follows: we first obtain an "Oracle" estimator
 by   minimized  a very tight upper bound of the Mean Square Error
 with changing the size of search window.
The "Oracle" estimator  depends on the unknown target function (original image),
whose concept  is developed in Donoho and Johnstone \cite{donoho1994ideal}.
So the "Oracle" estimator is not computable, but it can help us to find an available algorithm in mathematic theory.
We  second establish a theorem by the concept of the "Oracle" to show that  the Non-Local Means Filter can effectively deal with Poisson noise with some modification. Finally,  replacing the unknown target function by some estimators,
we construct our new algorithm called Non-Local Means Poisson Filter and demonstrate in statistic theory that the filter converges at the usual optimal rate. The filter is as simple as the classic Non-Local Means and the simulation results show that our filter is very competitive.

The remainder of this paper is organized as follow: we first introduce an "Oracle" estimator for Poisson noise based on the idea of Non-Local Means, and present a theorem to show the rate of convergence of the "Oracle" estimator in Section \ref{sec_main results}. We second construct an adaptive estimator according to the "Oracle" estimator and obtain some convergence theorems of the estimator in Section \ref{sec theory nlm}. Finally, we demonstrate in Section \ref{Sec:simulations} the ability of approach at restoring image contaminated by Poisson noise with a brief analysis.

\section{\label{sec_main results} The "Oracle" estimator}
\subsection{Some notations}
We
suppose that the original image of the object being photographed is a
integrable two-dimensional function $f(x)$, $x\in (0,1]\times(0,1]$. Let the
mean value of $f$ in a set $\mathbf{B}_x$ be
$$
\Lambda(\mathbf{B}_x)=N^2\int\limits_{\mathbf{B}_x}f(t)dt.  \label{Lam001}
$$
Typically we observe a discrete dataset of counts $\mathbf{Y}=\{\mathcal{N}(\mathbf{B}%
_x)$\}, where $\mathcal{N}(\mathbf{B}_x)$ is a Poisson random variable of intensity $%
\Lambda(\mathbf{B_x})$. We consider that if
$
\mathbf{B}_x \cap \mathbf{B}_y=\emptyset,
$
 then $\mathcal{N}(\mathbf{B}_x)$ is independent of $\mathcal{N}(\mathbf{B}_y)$. Suppose that $%
x=(x^{(1)},x^{(2)})\in \mathbf{I}=\{\frac{1}{N},\frac{1}{N},\cdots,1\}^2$,
and $\mathbf{B}_x=(x^{(1)}-1/N,x^{(1)}]\times(x^{(2)}-1/N,x^{(2)}]$. Then $\{%
\mathbf{B}_x\}_{x\in \mathbf{I}}$ is a partition of the square $(0,1]\times(0,1]$.
Using this partition  we get a discrete function $f(x)=\Lambda(\mathbf{B}_x)$, $%
x\in \mathbf{I}$. The denoising algorithm aims at estimating the underlying intensity
profile discrete function $f(x)=\Lambda(\mathbf{B}_x)$.
The image function $f$ is considered to be constant on each
 $\mathbf{B}_x$, $x\in \mathbf{I}$.
  Therefore $f(x)=\mathcal{N}(\mathbf{B}_x)$, $x\in \mathbf{I}$.
Furthermore, we can estimate the integrable function $p$ by the discrete
function $f$. Let
\begin{equation}
\label{Yx}
Y(x)=\mathcal{N}(\mathbf{B}_{x}),\,\,x\in \mathbf{I}.
\end{equation}%
This model has been used effectively in many contexts. The Poisson noise
model can be rewritten in the regression form
\begin{equation}
Y(x)=f(x)+\epsilon (x),\,\,x\in \mathbf{I,}  \label{Poisson noise model}
\end{equation}%
where $\epsilon (x)=\mathcal{N}(\mathbf{B}_{x})-f(x)$. It is easy to see that $\mathbb{E}(\epsilon (x))=0$ and $\mathbb{V}ar(\epsilon (x))=f(x)$.

Let us set some  notations to be used
throughout the paper. The Euclidean norm of a vector $x=\left(
x_{1},...,x_{d}\right) \in \mathbf{R}^{d}$ is denoted by $%
\left\Vert x\right\Vert _{2}=\left( \sum_{i=1}^{d}x_{i}^{2}\right)^{\frac{1}{2}} .$ The
supremum norm of $x$ is denoted by $\Vert x\Vert _{\infty }=\sup_{1\leq
i\leq d}\left\vert x_{i}\right\vert .$ The cardinality of a set $\mathbf{A}$ is denoted $\mathrm{card}\, \mathbf{A}$. For a positive integer $N$ the uniform $N\times N$ grid of pixels on the unit square is defined
by
\begin{equation}
\mathbf{I}=\left\{ \frac{1}{N},\frac{2}{N},\cdots ,\frac{N-1}{N},1\right\}
^{2}.  \label{def I}
\end{equation}%
Each element $x$ of the grid $\mathbf{I}$ will be called pixel. The number of pixels
is $n=N^{2}.$ For any pixel $x_{0}\in \mathbf{I}$ and a given $h>0,$ the
square window of pixels
\begin{equation}
\mathbf{U}_{x_{0},h}=\left\{ x\in \mathbf{I:\;}\Vert x-x_{0}\Vert _{\infty
}\leq h\right\}  \label{def search window}
\end{equation}%
will be called \emph{search window} at $x_{0}.$ We naturally take $h$ as a multiple of $\frac{1}{N}$ ($ h=\frac{k}{N}$ for some $k\in \{ 1, 2,\cdots,N\}$). The size of the square
search window $\mathbf{U}_{x_{0},h}$ is the positive integer number
\begin{equation}
M=(2Nh+1)^{2}=\mathrm{card\ }\mathbf{U}_{x_{0},h}. \label{defi M}
\end{equation}
 For any pixel $x\in \mathbf{U}%
_{x_{0},h}$ and a given $\eta >0$ a second square window of pixels
$
\mathbf{U}_{x,\eta }
$
will be called  \emph{ patch} at $x$. Like $h$, the parameter $\eta$ is also taken as a multiple of $\frac{1}{N}$. The size of the
 patch $\mathbf{U}_{x,\eta }$ is the positive integer
\begin{equation}
m=(2N\eta+1)^2=\mathrm{card\ }\mathbf{U}_{x_{0},\eta }.
\label{defi m}
\end{equation}

\subsection{The Non-Local Means algorithm}

The Non-Local Means algorithm (2005 \cite{buades2005review}) can be described as follows.
For any $x\in \mathbf{I}$,
\begin{equation}
\widetilde{f}=\sum_{x\in \mathbf{I}}w(x)Y(x),  \label{non local algorithm}
\end{equation}%
where the weights  $w\left(x\right) $ are given by
\begin{equation}
w(x)=e^{-\widetilde{\rho }_{x_0}^{2}(x)/H^{2}}\bigg/\sum_{x^{\prime }\in \mathbf{I}%
}e^{-\widetilde{\rho }_{x_0}^{2}(x^{\prime })/H^{2}},  \label{weights non local}
\end{equation}%
with
$$
\widetilde{\rho }_{x_0}^{2}=\sum_{y\in \mathbf{U}_{x_{0},\eta }}   \frac{\kappa
(y)|Y(y)-Y(T_x y)|^{2}}{\sum\limits_{y'\in \mathbf{U}_{x_{0},\eta }} \kappa(y')}.
$$%
Here $H$ is a bandwidth parameter, $\mathbf{U}_{x_{0},\eta }$ is given by (\ref{def search window}), $\kappa (y)>0$ are some fixed
kernel, and $T_x$ is the translation mapping:
\begin{equation}
T_x :x_{0}+y\rightarrow x+y
\label{difi Tx0x}
\end{equation}
 In practice the
bandwidth parameter $H$ is often taken as a linear function of $\sigma $
(see \cite{buades2005review}).

\subsection{Oracle estimator}

In order to adapt the Non-Local Means algorithm to the Poisson noise, we
introduce an "Oracle" estimator (for details on this concept see Donoho and
Johnstone (1994 \cite{donoho1994ideal})). Denote
\begin{equation}
f_{h}^{\ast }=\sum_{x\in \mathbf{U}_{x_0,h}}w_{h}^{\ast }Y(x),
\label{oracle_extimate}
\end{equation}%
where
\begin{equation}
w_{h}^{\ast }(x)=e^{-\frac{\rho_{f,x_0}^2(x)}{H^{2}(x_{0})}}  \bigg/\sum_{x^{\prime }\in \mathbf{U}_{x_0,h}}e^{-\frac{\rho_{f,x_0} ^{2}(x^{\prime })}{H^{2}(x_{0})}}  \label{oracle_weights}
\end{equation}%
with
\begin{equation}
\rho _{f,x_0}(x)\equiv |f(x)-f(x_{0})|,
\label{simulation function}
\end{equation}
  $H(x)$ is a control function subject to
\begin{equation}
\gamma=   \inf \{H(x):x\in \mathbf{I}\}
>0.
\label{condition of Hx}
\end{equation}
It is obvious that
\begin{equation}
\sum_{x\in \mathbf{U}_{x_0,h}}w_{h}^{\ast }(x)=1\quad \mathrm{and}\quad w_{h}^{\ast
}(x)\geq 0.
\label{condition of weights}
\end{equation}%
Note that the function $\rho _{f,x_0}(x) \geq 0$ characterizes the similarity of the image brightness
at the pixel $x$ with respect to the pixel $x_{0}$, therefore we shall call $%
\rho _{f,x_0}$ similarity function. The usual bias-variance decomposition  (cf. e.g. \citep{mandel1982use,terrell1992variable,fan1993local}) of the
Mean Squared Error (MSE)%
\begin{eqnarray}
\label{MSE}
&&\mathbb{E}\left(f(x_{0})-f_{h}^{\ast }(x_{0})\right) ^{2}
\\&=& \left(\sum_{x\in
\mathbf{U}_{x_0,h}}w_{h}^{\ast }(x)\left(f(x)-f(x_{0})\right) \right)
^{2}+\sum_{x\in \mathbf{U}_{x_0,h}}w_{h}^{\ast }(x)^{2}f(x) \nonumber\\
   &\leq& \left(\sum_{x\in
\mathbf{U}_{x_0,h}}w_{h}^{\ast }(x)|f(x)-f(x_{0})| \right)
^{2}+\sum_{x\in \mathbf{U}_{x_0,h}}w_{h}^{\ast }(x)^{2}f(x).\nonumber
\end{eqnarray}
The inequality (\ref{MSE}) combining with (\ref{simulation function}) implies the following upper bound
\begin{equation}
\mathbb{E}\left(f(x_{0})-f_{h}^{\ast }(x_{0})\right) ^{2}\leq g(w_{h}^{\ast
}(x)),  \label{upper_bound}
\end{equation}%
where
\begin{equation}
g(w)=\left(\sum_{x\in \mathbf{U}_{x_0,h}}w(x)\rho_{f,x_0}(x)\right) ^{2}+\sum_{x\in
\mathbf{U}_{x_0,h}}w(x)^{2}f(x).  \label{function upper bound}
\end{equation}%
We shall define a family of estimates by minimizing the function $g\left(w_{h}\right) $ by changing the width of the search window. With a Poisson noise in low-light conditions, the upper
bound of signal function is small, so we let $\Gamma =\sup \{f(x):x\in \mathbf{I}\}
$. According to the similarity phenomenon existing very often in natural
images, we suppose that the function $f$ satisfies the local H\"{o}lder
condition
\begin{equation}
|f(x)-f(y)|\leq L\Vert x-y\Vert _{\infty }^{\beta },\,\,\,\forall x,\,y\in
\mathbf{U}_{x_{0},h+\eta },  \label{Local Lip cond}
\end{equation}%
where $\beta >0$ and $L>0$ are constants,  $h>0$, $\eta >0$ and $x_{0}\in \mathbf{I%
}.$ The following theorem gives the rate of convergence of the "Oracle"
estimator and the proper width $h$ of the search window.

\begin{theorem}
\label{th_oracle} Assume that $h=\left(\frac{\Gamma%
}{4\beta L^2}\right)^{\frac{1}{2\beta+2}}n^{-\frac{1}{2\beta+2}}$ and $\gamma>\sqrt
2 Lh^{\beta}$. Suppose that the function $f$ satisfies the local H\"{o}%
lder condition (\ref{Local Lip cond}) and $f^*_h(x_0)$ be given by (\ref{oracle_extimate}). Then
\begin{equation}
\mathbb{E}\left(f^*_h(x_0)-f(x_0)\right)^2\leq c_0 n^{-\frac{2\beta}{%
2\beta+2}},  \label{rate_oracle}
\end{equation}
where
\begin{equation}
c_0=\frac{2^{\frac{2\beta + 6}{2\beta +2}} \Gamma^{\frac{2\beta}{2\beta +2}}
L^{\frac{4}{2\beta +2}} }{\beta^{\frac{2\beta}{2\beta +2}}}.
\end{equation}
\end{theorem}

For the proof of this theorem see Section \ref{proof of th oracle}.

This theorem shows that at least from the practical point of view, it is
justified to optimize the upper bound $g(w)$ instead of optimizing the risk $%
\mathbb{E}\left(f_{h}^{\ast }(x_{0})-f(x_{0})\right) ^{2}$ itself.
The theorem also justifies that we can choose a small search window in place
of the whole observed image to estimate a point, without loss of visual
quality. That is why we only consider small search windows for the
simulations of our algorithm.

\section{Non-Local Means Poisson Filter \label{sec theory nlm} }
\subsection{Construction of Non-Local Means Poisson Filter}
With the theory of "Oracle" estimator, we construct the Non-Local Means Poisson Filter. Let $h>0$ and $\eta >0$ be fixed numbers.
Since $%
|f(x)-f(x_{0})|^{2}=\mathbb{E}|Y(x)-Y(x_{0})|^{2}-(f(x_0)+f(x))$, an obvious
estimator of $\mathbb{E}\left\vert Y(x)-Y(x_{0})\right\vert ^{2}$ is given by
$$
\frac{1}{M}\sum_{y\in {\mathbf{U}_{x_{0},\eta }}%
}|Y(y)-Y(T_x y)|^{2},
$$
where $T_x$ is given by (\ref{difi Tx0x}), and $(f(x_0)+f(x))$ is estimated by $2\overline{f}(x_0)$, where
$$
\overline{f}(x_0)=\frac{1}{M}\sum_{x\in \mathbf{U}_{x_0,h}}f(x).
$$
Define an estimated similarity function $\widehat{\rho }_{x_0}$ by
\begin{equation}
\widehat{\rho }_{x_0}^{2}(x)= \left(\frac{1}{M}\sum_{y\in {\mathbf{U}_{x_{0},\eta }}%
}|Y(y)-Y(T_x y)|^{2}-2\overline{f}(x_0)\right)^+.  \label{estimator similar1}
\end{equation}%

The following theorem implies that it is reasonable to let  $\widehat{\rho}_{x_0}(x)$  be the estimator
of $\rho_{f,x_0}(x)$.

\begin{theorem}
\label{th similar function} Assume that $h=\left(\frac{\Gamma%
}{4\beta L^2}\right)^{\frac{1}{2\beta+2}}n^{-\frac{1}{2\beta+2}}$ and $\eta
=c_{1}n^{-\alpha }$ $\left(\frac{(1-\beta)^{+}}{2\beta +2}<\alpha <\frac{1%
}{2}\right) $.  Suppose that the function $f$ satisfies the
local H\"{o}lder condition (\ref{Local Lip cond}) and $\widehat{\rho }%
_{x_0}^{2}(x)$ is given by (\ref{estimator similar1}). Then there is a constant $c_{2}$ such that
\begin{equation}
\mathbb{P}\left\{ \max_{x\in \mathbf{U}_{x_{0},h}}\left\vert \widehat{\rho }%
_{x_0}^{2}(x)-\rho_{f,x_0}^2(x)\right\vert \geq c_{2}n^{\alpha -\frac{1}{2}}\sqrt{\ln n}%
\right\} \leq O\left(n^{-1}\right) .  \label{rate similar function}
\end{equation}
\end{theorem}

For the proof of this theorem see Section \ref{proof of similar condition}.

 As a result, it is natural to define an adaptive estimator  $\widehat{f}_{h }$ by
\begin{equation}
\widehat{f}_{h }(x_{0})=\sum_{x\in  \mathbf{U}_{x_{0},h}}\widehat{w}%
_{h}(x)Y(x),  \label{estimate001}
\end{equation}%
where \begin{equation}
\widehat{w}_{h}=e^{-\frac{\widehat{\rho }_{x_0}^{2}(x)}{H^{2}}}\bigg/\sum_{x^{\prime }\in
\mathbf{U}_{x_{0},h}}e^{-\frac{\widehat{\rho }_{x_0}^{2}(x^{\prime })}{%
H^{2}}}.  \label{estimate_weights}
\end{equation}
 and
  $\mathbf{U}_{x_{0},h}$ given by (\ref{def search window}).

\subsection{Convergence theorem of Non-Local Means Poisson Filter}
Now, we turn to the study of the convergence of the Non-Local Means Poisson Filter. Due to the difficulty in dealing with the dependence of the weights we shall consider a slightly modified version of the proposed algorithm:  we divide  the set of pixels  into two independent parts,  so that the weights are constructed from the one part,  and  the estimation of the target function is a weighted mean  along  the other part. More precisely,   we split the set of pixels into two parts $\mathbf{I}=\mathbf{I}'_{x_{0}}\cup \mathbf{I}''_{x_{0}}$ for any $x_0\in \mathbf{I}$ where
$$
\mathbf{I}'_{x_{0}}=\left\{ x_{0}+\left( \frac{i}{N},\frac{j}{N}\right) \in
\mathbf{I}:i+j \mathrm{\, is\, even} \right\} ,
$$%
and $\mathbf{I}''_{x_{0}}=\mathbf{I}\diagdown \mathbf{I}'_{x_{0}}.$

Define an estimated similarity function $\widehat{\rho }_{x_0}$ by
\begin{equation}
\widehat{\rho }_{x_0}^{'2}(x)= \left(\frac{1}{\mathrm{card}%
\mathbf{U}''_{x_{0},\eta }}\sum_{y\in {\mathbf{U}''_{x_{0},\eta }}%
}|Y(y)-Y(T_x y)|^{2}-2\overline{f}'(x_0)\right)^+,  \quad x\in \mathbf{U}'_{x_0,h}\label{estimator similar}
\end{equation}%
where
\begin{equation}
\overline{f}'(x)=\frac{1}{\mathrm{card} \mathbf{U}''_{x_0,h}} \sum_{y\in
\mathbf{U}''_{x_0,h}}Y(y),
\end{equation}
and $\mathbf{U}_{x_{0},\eta}^{''}=\mathbf{U}_{x_{0},\eta}\cap \mathbf{I}''
_{x_{0}}$ with $\mathbf{U}_{x_{0},h}$ given by (\ref{def search window}).
The adaptive estimator $\widehat{f}'_{h }$ is denoted by
\begin{equation}
\widehat{f}'_{h }(x_{0})=\sum_{x\in \mathbf{U}_{x_{0},h}^{\prime }}\widehat{w}'%
_h(x)Y(x),  \label{estimate002}
\end{equation}%
where  $\mathbf{U}_{x_{0},h}^{\prime }=\mathbf{U}_{x_{0},h}\cap \mathbf{I}'_{x_{0}}$
 and
 \begin{equation}
\widehat{w}'_h=e^{-\frac{\widehat{\rho }_{x_0}^{'2}(x)}{H^2(X_0)}}\bigg/\sum_{x^{\prime }\in \mathbf{U}_{x_{0},h}^{\prime }}e^{-%
\frac{\widehat{\rho }_{x_0}^{'2}(x^{\prime })}{H^2(x_0)}}.  \label{estimate_weights002}
\end{equation}

In the next theorem we prove that the Mean Squared Error of the estimator $%
\widehat{f}'_{h}(x_{0})$ converges at the rate $n^{-\frac{2\beta }{2\beta +2}%
} $ which is the usual optimal rate of convergence for a given H\"{o}lder
smoothness $\beta >0$ (see e.g. Fan and Gijbels (1996 \cite{FanGijbels1996}%
)).

\begin{theorem}
\label{th rate estimator} Let $\eta=c_3n^{-\alpha}$, $h=\left(%
\frac{\Gamma}{4\beta L^2}\right)^{\frac{1}{2\beta+2}}n^{-\frac{1}{2\beta+2}}$, $H(x_0)>4c_2n^{\alpha-\frac{1}{2}}\sqrt{\ln n}$  and $\gamma>\max\{\sqrt
2 Lh^{\beta},4c_2n^{\alpha-\frac{1}{2}}\sqrt{\ln n}\}$. Suppose that the function f satisfies the H\"{o}%
lder condition (\ref{Local Lip cond}) and $\widehat{f}'_{h}(x_0)$ is given by (\ref{estimator similar}).  Then
\begin{equation}
\mathbb{E} \left(\widehat{f}'_{h}(x_0)-f(x_0)\right) ^2 \leq c_4n^{-%
\frac{2\beta}{2\beta+2}},  \label{rate etimator}
\end{equation}
where
$$
c_4= 8\left(\frac{2^{\frac{2\beta + 6}{2\beta +2}} \Gamma^{\frac{2\beta}{%
2\beta +2}} L^{\frac{4}{2\beta +2}} }{\beta^{\frac{2\beta}{2\beta +2}}}
\right) ^2.
$$
\end{theorem}

For the proof of this theorem see Section \ref{proof of similar condition}.

\section{\label{Sec:simulations}Simulation results}

\subsection{Computational algorithm}

Throughout the simulations, we use the following algorithm to compute the
Non-Local Means Poisson estimator $\widehat{f}_{h}(x_{0}).$ The input values of the
algorithm are $Y\left(x\right) ,$ $x\in \mathbf{I}$ (the image),  and two numbers $m $ and $M $  are given by (\ref{defi m}) and (\ref{defi M}) respectively. In order to improve
the results, we introduce a smoothed version of the estimated similarity
distance%
\begin{equation}
\widehat{\rho }_{\kappa,x_0}^{2}(x)=\left(\sum_{y\in \mathbf{U}_{x_{0},\eta }}  \frac{\kappa \left(y\right)
|Y(y)-Y(T_x y)|^{2}-2\overline{f}(x_{0})}{\sum\limits_{y'\in \mathbf{U}_{x_{0},\eta }} \kappa(y') }\right) ^{+}.
\label{empir simil func}
\end{equation}%
As smoothing kernels $\kappa$ we use the Gaussian kernel
\begin{equation}
\kappa_{g}(y,h_{g})=\exp \left(-\frac{N^{2}\Vert y-x_{0}\Vert _{2}^{2}}{2h_{g}}%
\right) ,  \label{s4kg}
\end{equation}%
where $h_{g}$ is a constant, and the following kernel: for $y \in \mathbf{U}_{x_0,\eta}$,
\begin{equation}
\kappa_{0}\left( y\right) =\sum_{k=\max(1,j)}^{N\eta}\frac{1}{(2k+1)^2}, \label{def k0}
\end{equation}%
if $\|y-x_0\|_{\infty}=\frac{j}{N}$ for some $j\in \{0,1,\cdots,N\eta\}$. We shall also use the
rectangular kernel%
\begin{equation}
\kappa_{r}\left(y\right) =\left\{
\begin{array}{ll}
\frac{1}{\mathrm{card}\mathbf{U}_{x_{0},\eta }}, & y\in \mathbf{U}%
_{x_{0},\eta }, \\
0, & \mathrm{otherwise.}%
\end{array}%
\right.   \label{rect kernel}
\end{equation}%
For the simulation we use the kernel $\kappa_{0}(y)$ defined by (\ref{def k0}). We
have seen experimentally that when we take the filtering function $%
H^{2}(x_{0})$ as $\mu \cdot \sqrt{\overline{f}(x_{0})}$, where $\mu $ is a
constant depending on the character of the image, to obtain a denoising of
high visual quality. We mention that throughout the paper we symmetrize
images near the boundary.

\noindent\rule{\textwidth}{.2pt}

\textbf{Algorithm}\quad Non-Local Means Poisson Filter (NLMPF)

\noindent\rule{\textwidth}{.2pt}

Let $\{M,m,h_g\}$ be the parameters.

Repeat for each $x_0\in \mathbf{I}$

\quad - compute

\quad\textbf{ Step 1}

\quad \quad $\widehat{w}(x)=exp(-\widehat{\rho }_{\kappa,x_0}^2(x)/H^2(x_0))$

\quad \quad $\widehat{f}_1(x_0)=\sum_{x\in \mathbf{U}_{x_0,h}}\widehat{w}(x)Y(x)\big/%
\sum_{x\in \mathbf{U}_{x_0,h}}w(x)$

\quad \textbf{Step 2}

\quad \quad If $\frac{1}{(2d+1)^1}\sum _{\|x-x_0\|\leq d/N} \widehat{f}_1(x)<\delta$

\quad \quad compute $\widehat{f}(x_{0})= \sum _{\|x-x_0\|\leq d/N}
\kappa_{g}(x,h_g)\widehat{f}_1(x)/\sum _{\|x-x_0\|\leq d/N} \kappa_{g}(x,h_g)$

\quad \quad else $\widehat{f}(x_{0})=\widehat{f}_1(x_0).$

\noindent\rule{\textwidth}{.2pt}
{\em Note:} we take $\delta=15$.

\begin{center}
\begin{figure}[tbp]
\renewcommand{\arraystretch}{0.5} \addtolength{\tabcolsep}{-6pt} \vskip3mm {%
\fontsize{8pt}{\baselineskip}\selectfont
\begin{tabular}{ccccc}
\includegraphics[width=0.20\linewidth]{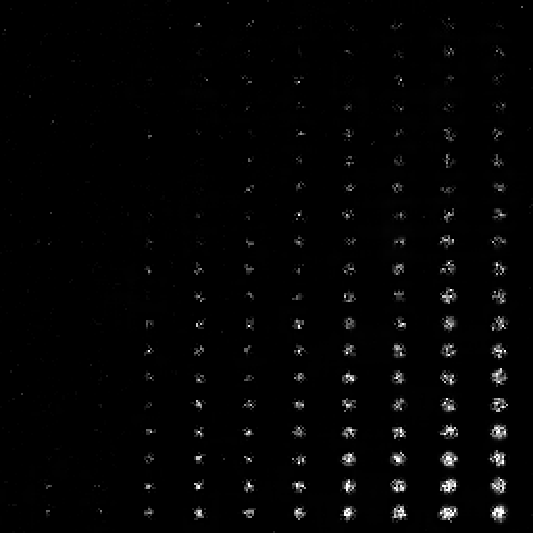} & %
\includegraphics[width=0.20\linewidth]{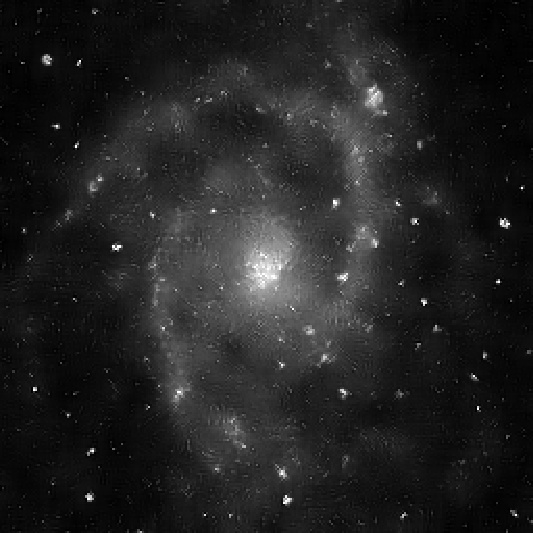} & %
\includegraphics[width=0.20\linewidth]{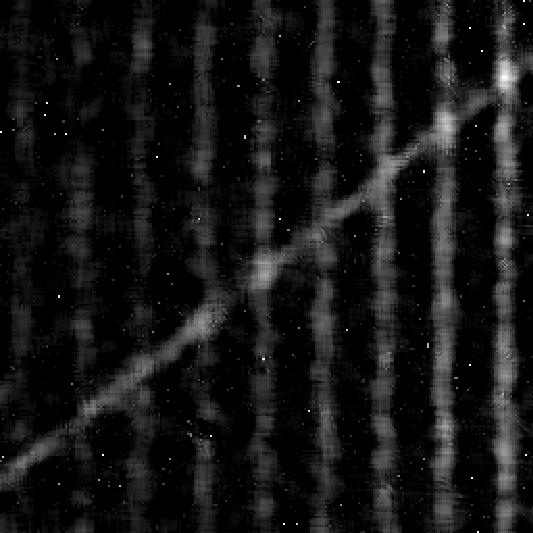} & %
\includegraphics[width=0.20\linewidth]{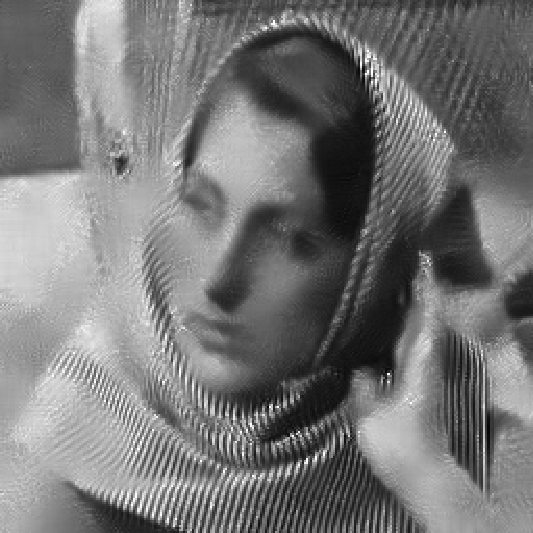} & %
\includegraphics[width=0.20\linewidth]{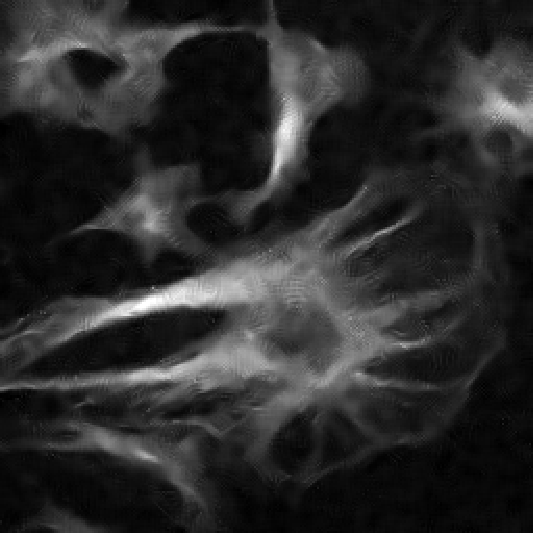} \\
NMISE=0.1009 & NMISE=0.0593 & NMISE=0.0368 & NMISE=0.1150 & NMISE=0.0870 \\
$M=19\times19$ & $M=13\times13$ & $M=9\times9$ & $M=15\times15$ & $%
M=7\times7 $ \\
$m=13\times13$ & $m=3\times3$ & $m=21\times21$ & $m=21\times21$ & $%
m=13\times13$ \\
(a) Spots & (b) Galaxy & (c) Ridges & (d) Barbara & (e) Cells%
\end{tabular}
}
\caption{{\protect\small These images restored by the first step of our
algorithm. }}
\label{Fig step1}
\end{figure}
\end{center}

\subsection{Numerical performance of the Non-Local Means Poisson Filter}

By simulations we found that the images with brightness between $0$ and $255$
(like Barbara) are well denoised by the first step, but for the low count
levels images (with brightness less than $\mu$), the restored images by NLMPF are not smooth enough (see Figure %
\ref{Fig step1}). This explains why for the low count level images, we
smooth the restored images by step 2.

Our experiments are done in the same way as in \cite{ZHANG2008WAVELETS} and %
\cite{MAKITALO2009INVERSION} to produce comparable results; we also use the
same set of test images (all of $256\times 256$ in size): Spots $[0.08,4.99]$%
, Galaxy $[0,5]$, Ridges $[0.05,0.85]$, Barbara $[0.93,15.73]$, and Cells $%
[0.53,16.93]$. The authors of \cite{ZHANG2008WAVELETS} and %
\cite{MAKITALO2009INVERSION} kindly provided us with their programs and the
test images. A matlab implementation of the algorithms derived in
this paper is available online\footnote{http://www.pami.sjtu.edu.cn/people/jinqy/}. This unoptimized implemen-tation processes the set of $256 \times 256$ test images   $145$  seconds with  a search window of size $15 \times 15$  and patches
of size $21\times 21$, $52$  seconds with  a search window of size $9 \times 9$  and patches
of size $21\times 21$.
 The
computational time is of about 10s per iteration on a $256 \times 256$
image and Matlab on an Intel Pentium Dual CPU T3200 32-bit @ 2.00GHz
CPU 3.00GHz.

\begin{figure}[tbp]
\renewcommand{\arraystretch}{0.5} \addtolength{\tabcolsep}{-5pt} \vskip3mm {%
\fontsize{8pt}{\baselineskip}\selectfont
\begin{tabular}{ccc}
\includegraphics[width=0.33\linewidth]{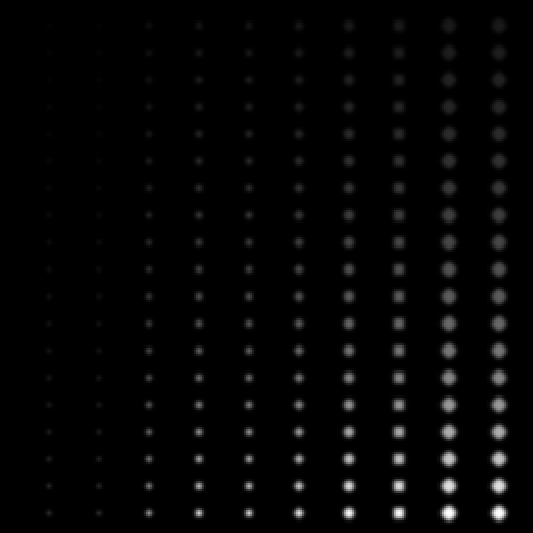} & %
\includegraphics[width=0.33\linewidth]{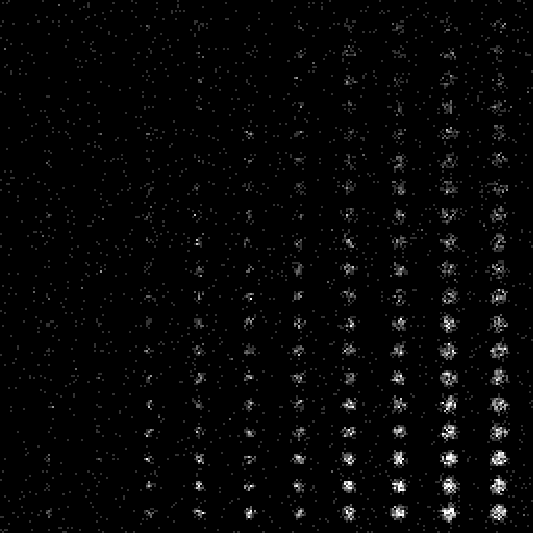}&
\includegraphics[width=0.33\linewidth]{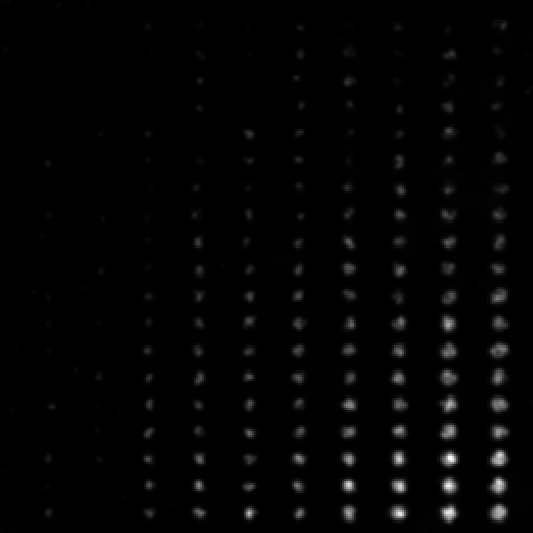}\\
(a) Original image & (b) Noisy image&
 (c) NLMPF \\&&$NMISE = 0.0260$ \\&&$PSNR = 31.45$ \\
\includegraphics[width=0.33\linewidth]{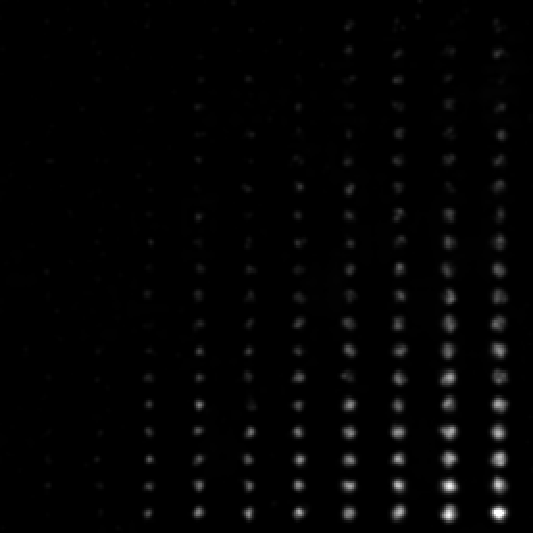} &
\includegraphics[width=0.33\linewidth]{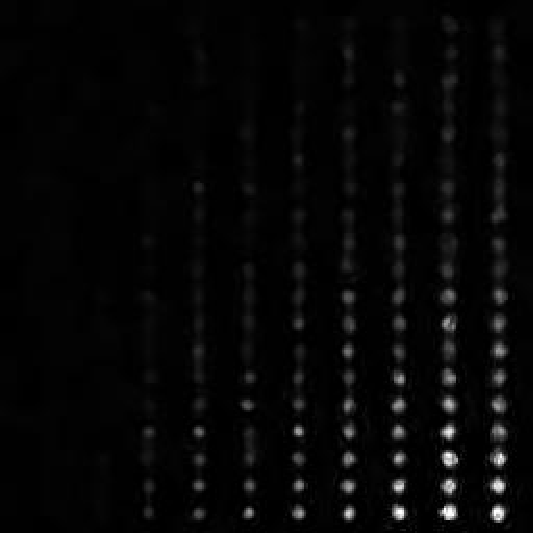} &
\includegraphics[width=0.33\linewidth]{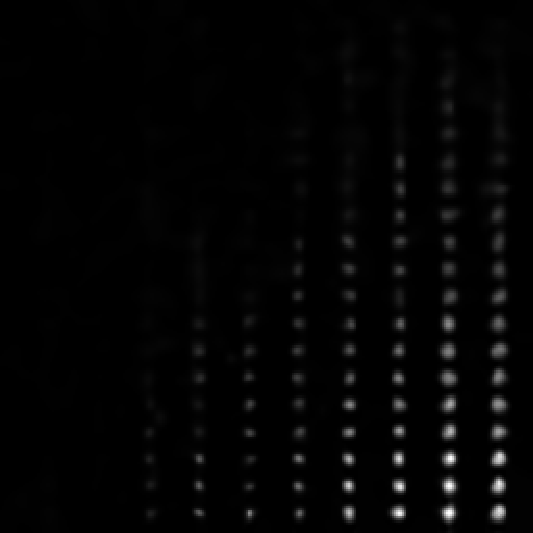}\\
(d)OWPNF& (e) Poisson NLM & (f) EUI+BM3D\\
 $NMISE=0.0259$&$NMISE=0.0790$&$NMISE=0.0358$\\
 $PSNR=31.31$&$PSNR=31.12$&$PSNR=31.95$\\
\includegraphics[width=0.33\linewidth]{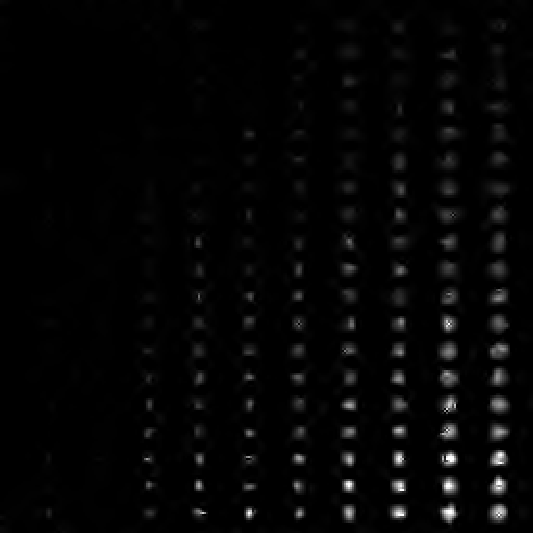} &
\includegraphics[width=0.33\linewidth]{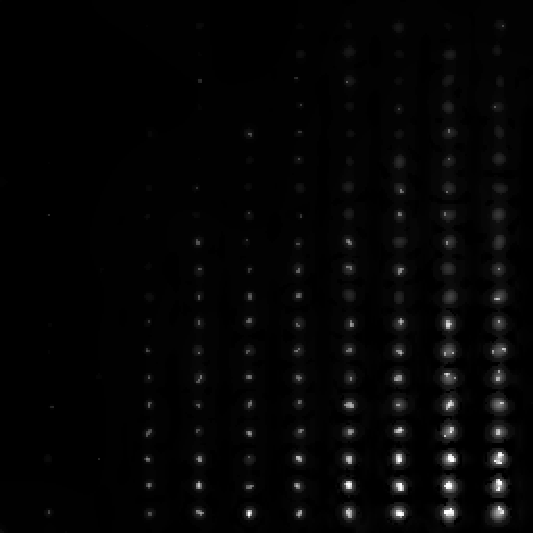}   \\
 (g)  MS-VST + $7/9$& (h) MS-VST + B3 \\
  $NMISE = 0.0602$&$NMISE = 0.0810$ \\
  $PSNR = 31.64$&$PSNR = 31.35$
\end{tabular}
}
\caption{{\protect\small Denoising an image of simulated spots of different
radii (image size: $256 \times 256$).
(a) simulated sources (amplitudes $\in
[0.08, 4.99]$; background $= 0.03$);
(b) observed counts;
(c) NLMPF ($M=19\times19$, $m=13\times13$, $d=3$, $\protect\sigma_H=2.5 $%
, $\protect\mu=1$ $NMISE = 0.0260$);
(d) Optimal
Weights Filter ($M=19\times19$, $m=13\times13$, $d=2$ and $H=1, $ $NMISE =
0.0259$);
(e) Poisson NLM ($%
NMISE=0.0790$);
(f) Exact unbiased inverse + BM3D ($%
NMISE=0.0358$) ;
(g) MS-VST + $7/9$ biorthogonal wavelet ($J = 5$, $FPR = 0.01$%
,$N_{max}= 5$ iterations, $NMISE = 0.0602$);
(h) MS-VST + B3 isotropic
wavelet ($J = 5$, $FPR = 0.01$, $N_{max}= 5$ iterations, $NMISE = 0.0810$).}}
\label{Fig spots}
\end{figure}

\begin{figure}[tbp]
\renewcommand{\arraystretch}{0.5} \addtolength{\tabcolsep}{-5pt} \vskip3mm {%
\fontsize{8pt}{\baselineskip}\selectfont
\begin{tabular}{ccc}
\includegraphics[width=0.33\linewidth]{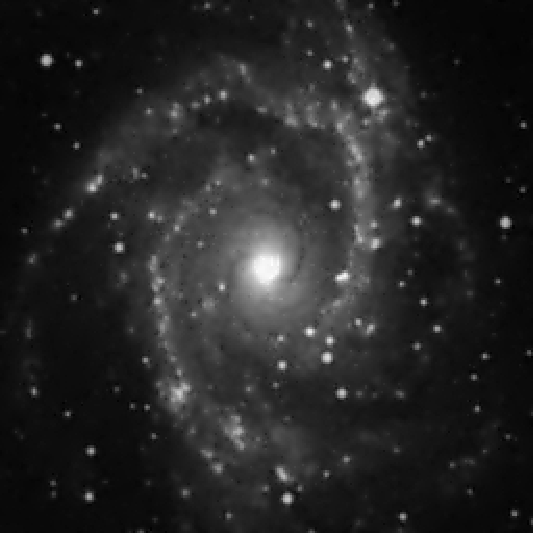} & %
\includegraphics[width=0.33\linewidth]{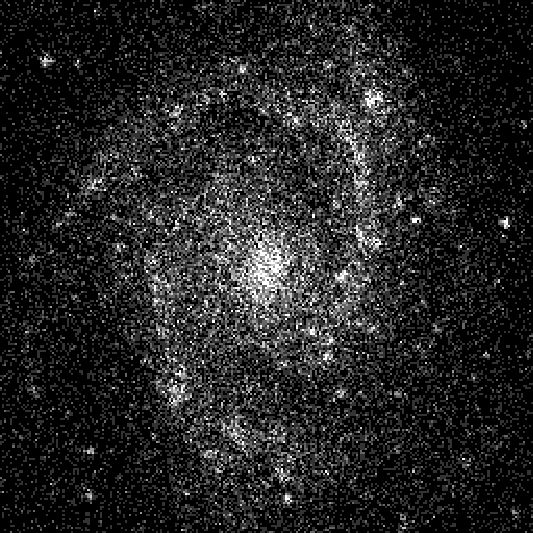} &
\includegraphics[width=0.33\linewidth]{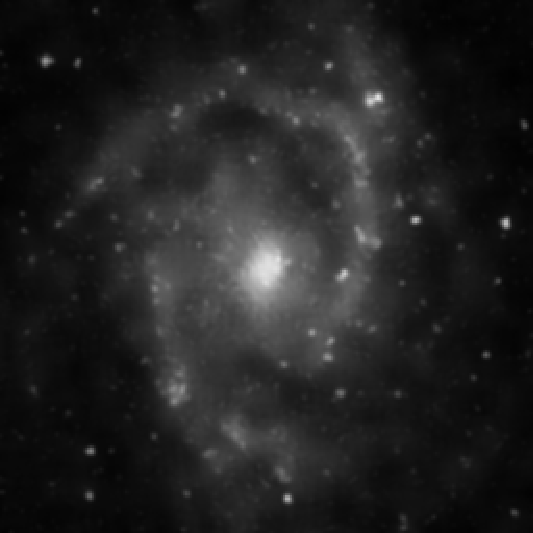}    \\
(a) Original image & (b) Noisy image &(c) NLMPF\\&& $NMISE = 0.0284$\\&& $PSNR = 28.09$\\
\includegraphics[width=0.33\linewidth]{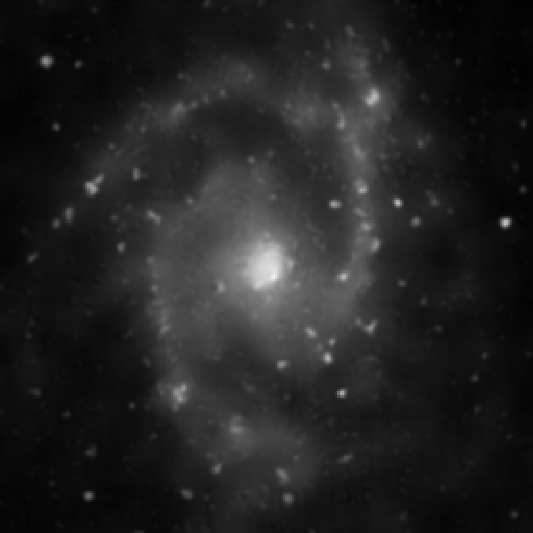} &
\includegraphics[width=0.33\linewidth]{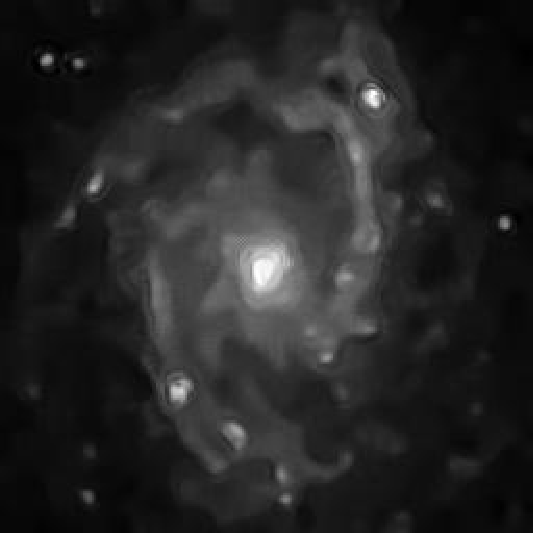} &
\includegraphics[width=0.33\linewidth]{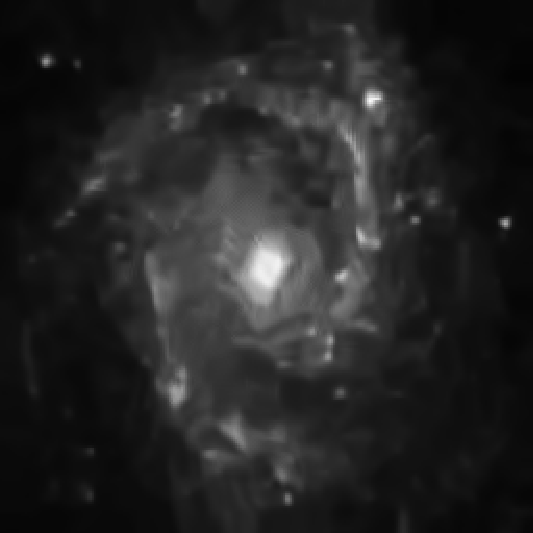} \\
(d)OWPNF& (e) Poisson NLM & (f) EUI+BM3D\\
 $NMISE=0.0285$&$NMISE=0.0346$&$NMISE=0.0297$\\
 $PSNR=27.80$&$PSNR=27.77$&$PSNR=28.04$\\
\includegraphics[width=0.33\linewidth]{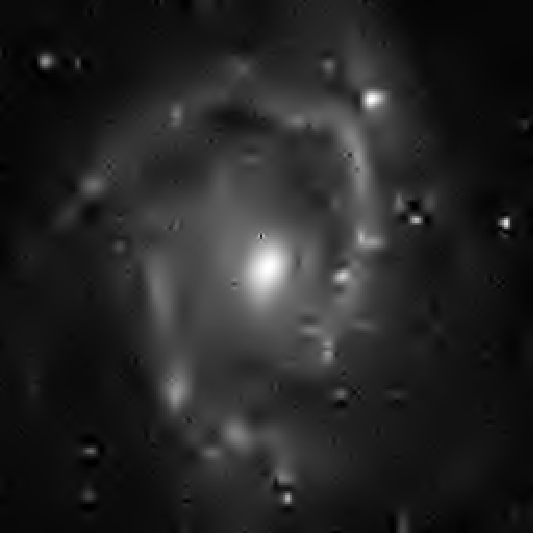} & %
\includegraphics[width=0.33\linewidth]{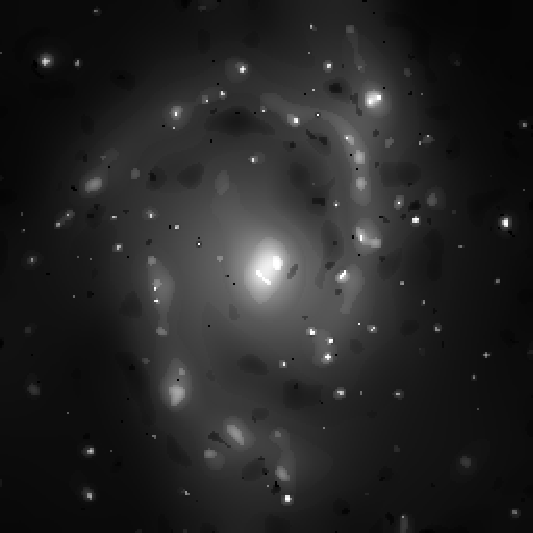}   \\
 (g)  MS-VST + $7/9$& (h) MS-VST + B3 \\
  $NMISE = 0.0357$&$NMISE = 0.0338$\\
  $PSNR = 27.57$&$PSNR = 27.94$
\end{tabular}
}
\caption{{\protect\small Denoising a galaxy image (image size: $256 \times
256$).
(a) galaxy image (intensity $\in [0, 5]$);
(b) observed counts;
(c)
NLMPF ($M=13\times13$, $m=3\times3$, $d=2$, $\protect\sigma%
_H=1$, $\protect\mu=0.6$ $NMISE = 0.0284$);
(d)
Optimal Weights Filter ($M=15\times15$, $m=5\times5$, $d=2$ and $H=1$, $%
NMISE = 0.0285$);
(e) Poisson NLM ($%
NMISE=0.0346$);
(f) Exact unbiased inverse +
BM3D ($NMISE=0.0297$) ;
(g) MS-VST + $7/9$ biorthogonal wavelet ($J = 5$, $FPR
= 0.0001$, $N_{max}= 5$ iterations, $NMISE = 0.0357$);
(h) MS-VST + B3
isotropic wavelet ($J = 3$, $FPR = 0.0001$, $N_{max}= 10$ iterations, $NMISE
= 0.0338$).}}
\label{Fig Galaxy}
\end{figure}

\begin{figure}[tbp]
\renewcommand{\arraystretch}{0.5} \addtolength{\tabcolsep}{-5pt} \vskip3mm {%
\fontsize{8pt}{\baselineskip}\selectfont
\begin{center}
\begin{tabular}{ccc}
\includegraphics[width=0.33\linewidth]{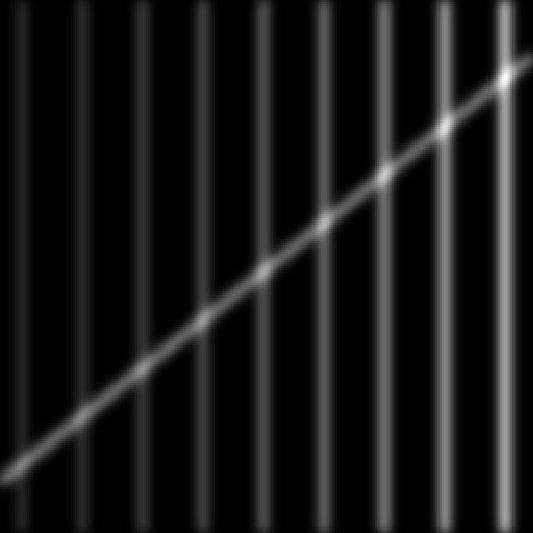} & %
\includegraphics[width=0.33\linewidth]{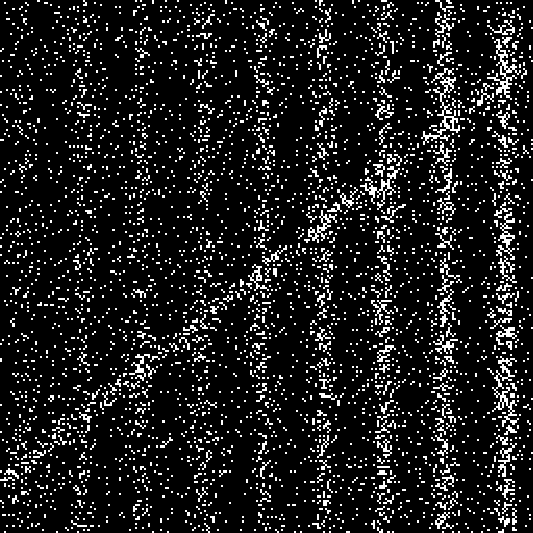} &
\includegraphics[width=0.33\linewidth]{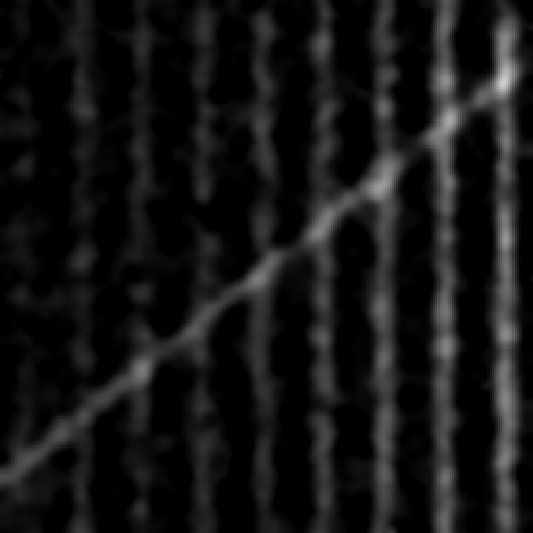}  \\
(a) Original image & (b) Noisy image&
 (c) NLMPF \\&&$NMISE = 0.0140$ \\&&$PSNR = 24.69$ \\
\includegraphics[width=0.33\linewidth]{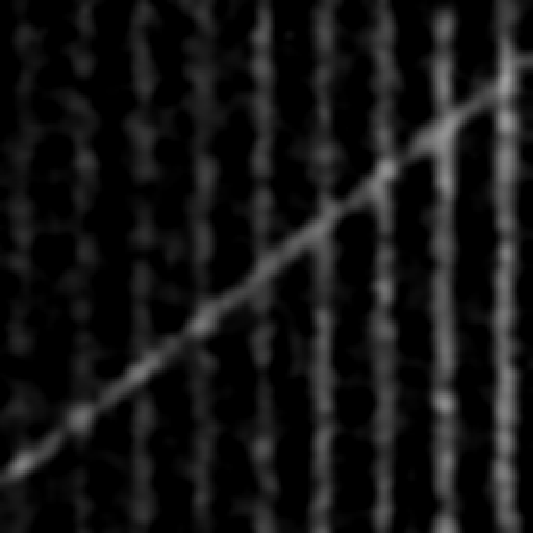} &
\includegraphics[width=0.33\linewidth]{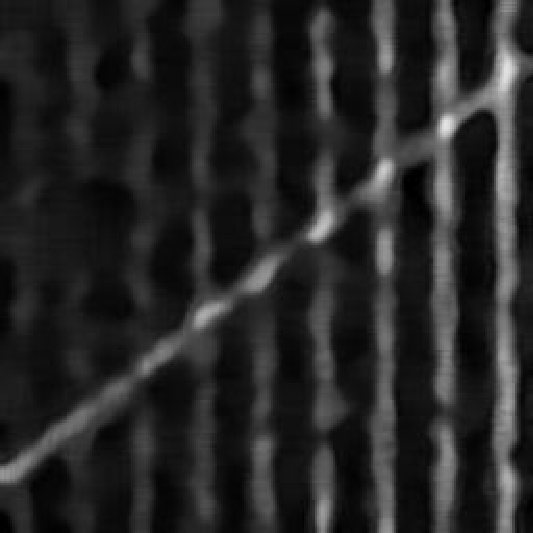} &
\includegraphics[width=0.33\linewidth]{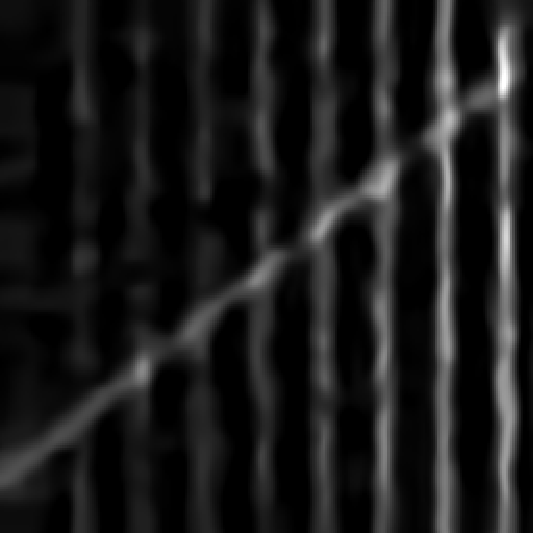} \\
(d)OWPNF& (e) Poisson NLM & (f) EUI+BM3D\\
 $NMISE=0.0.0162$&$NMISE=0.0154$&$NMISE=0.0.0121$\\
 $PSNR=23.90$&$PSNR=24.94$&$PSNR=25.89$\\
\includegraphics[width=0.33\linewidth]{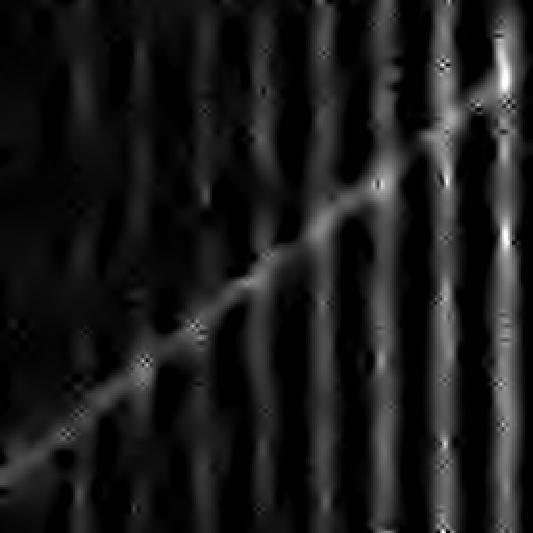} & %
\includegraphics[width=0.33\linewidth]{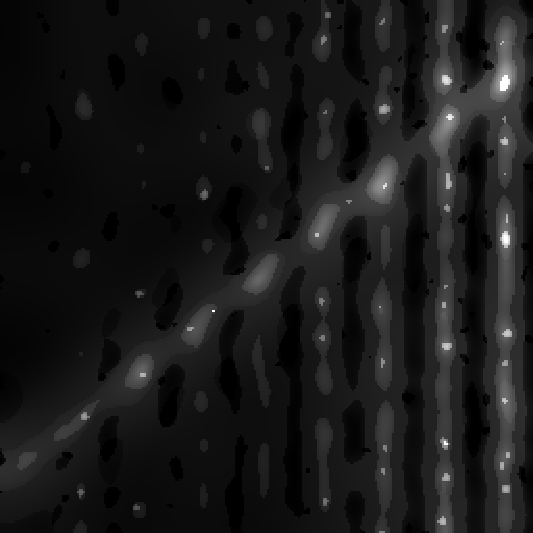}   \\
 (g)  MS-VST + $7/9$& (h) MS-VST + B3 \\
  $NMISE = 0.0.0193$&$NMISE = 0.0416$\\
  $PSNR = 24.49$&$PSNR = 24.03$
\end{tabular}
\end{center}
}
\caption{{\protect\small Poisson denoising of smooth ridges (image size: $%
256 \times 256$). (a) intensity image (the peak intensities of the $9$
vertical ridges vary progressively from $0.1$ to $0.5$; the inclined ridge
has a maximum intensity of $0.3$; background $= 0.05$); (b) Poisson noisy
image; (c) NLMPF ($M=9\times9$, $m=21\times21$, $d=4$, $%
\protect\sigma_H=0.5$, $\protect\mu=0.4$, $NMISE = 0.0140$);
(d) Optimal Weights Filter ($M=9\times9$, $m=19\times19$, $d=3$ and $%
H=2$, $NMISE = 0.0162$);
(e) Poisson NLM ($%
NMISE=0.0154$);
 (f) Exact
unbiased inverse + BM3D ($NMISE=0.0121$);
(g) MS-VST + $%
7/9 $ biorthogonal wavelet ($J = 5$, $FPR = 0.001$, $N_{max}= 5$ iterations,
$NMISE = 0.0193$);
(h) MS-VST + B3 isotropic wavelet ($J = 3$, $FPR = 0.00001
$, $N_{max}= 10$ iterations, $NMISE = 0.0416$).}}
\label{Fig Ridges}
\end{figure}

\begin{figure}[tbp]
\renewcommand{\arraystretch}{0.5} \addtolength{\tabcolsep}{-5pt} \vskip3mm {%
\fontsize{8pt}{\baselineskip}\selectfont
\begin{tabular}{ccc}
\includegraphics[width=0.33\linewidth]{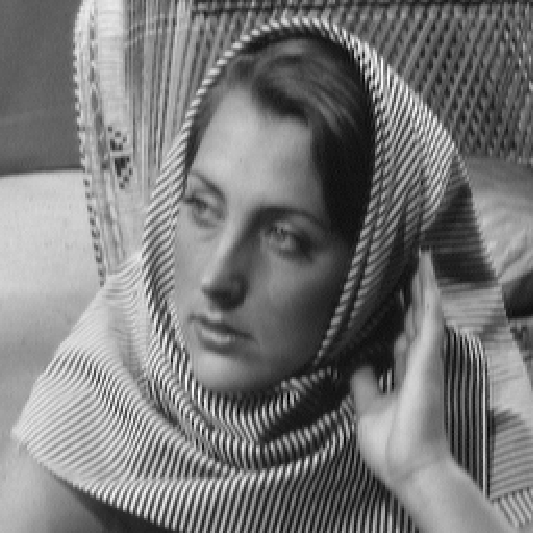} & %
\includegraphics[width=0.33\linewidth]{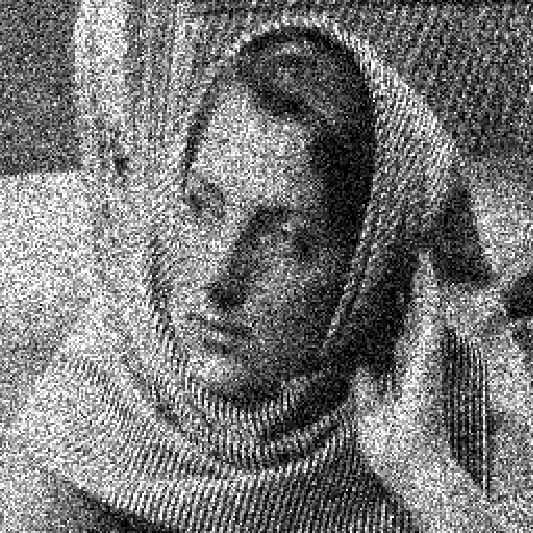} &
\includegraphics[width=0.33\linewidth]{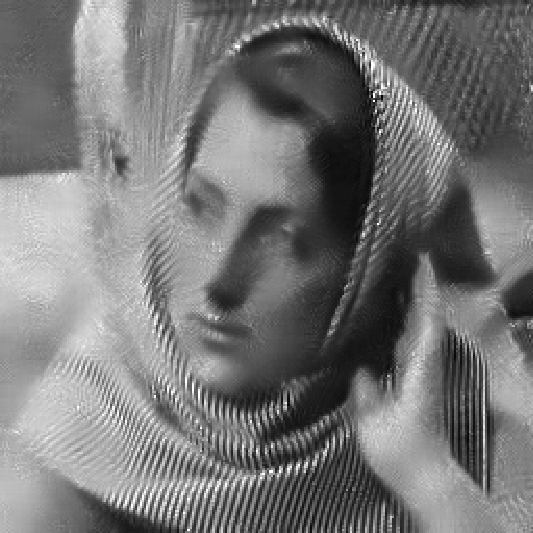}    \\
(a) Original image & (b) Noisy image&
 (c) NLMPF \\&&$NMISE = 0.1150$ \\&&$PSNR = 24.71$ \\
\includegraphics[width=0.33\linewidth]{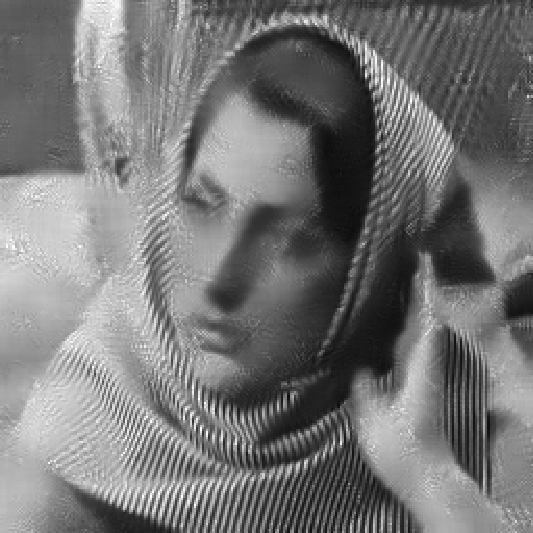} &
\includegraphics[width=0.33\linewidth]{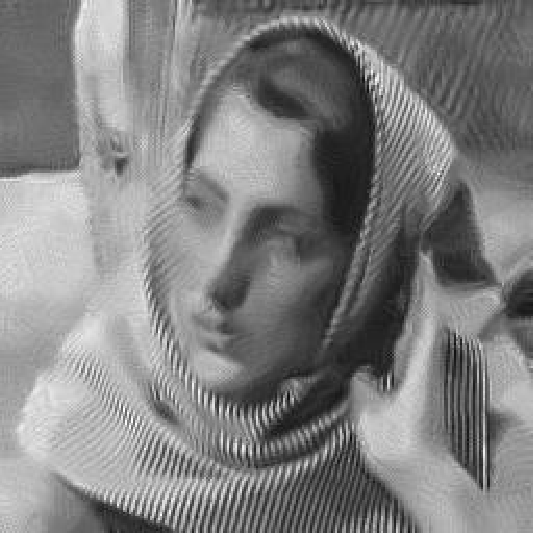} &
\includegraphics[width=0.33\linewidth]{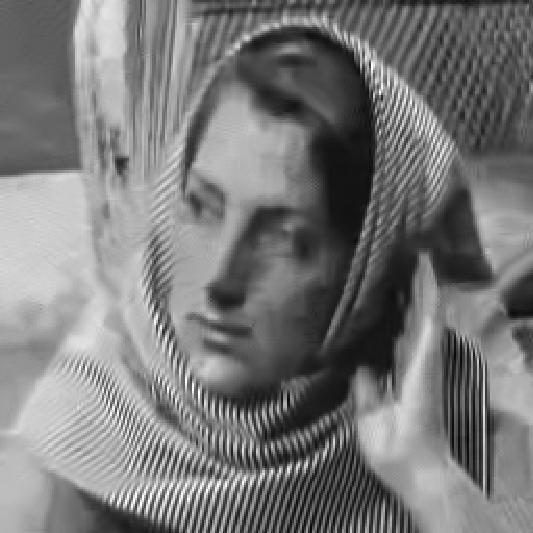} \\
(d)OWPNF& (e) Poisson NLM & (f) EUI+BM3D\\
 $NMISE=0.1061$&$NMISE=0.1207$&$NMISE=0.0683$\\
 $PSNR=24.60$&$PSNR=24.72$&$PSNR=25.92$\\
\includegraphics[width=0.33\linewidth]{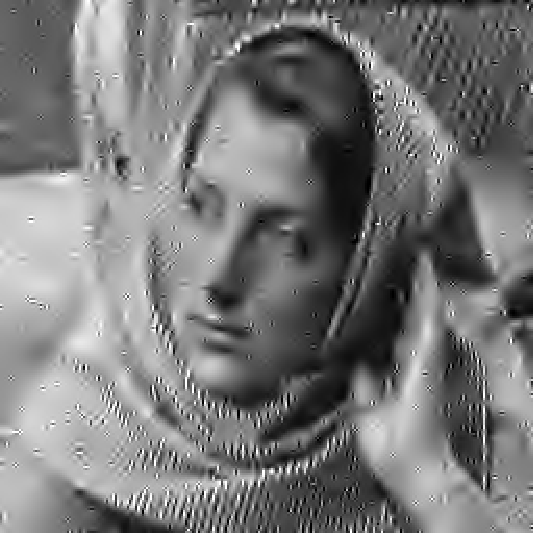} & %
\includegraphics[width=0.33\linewidth]{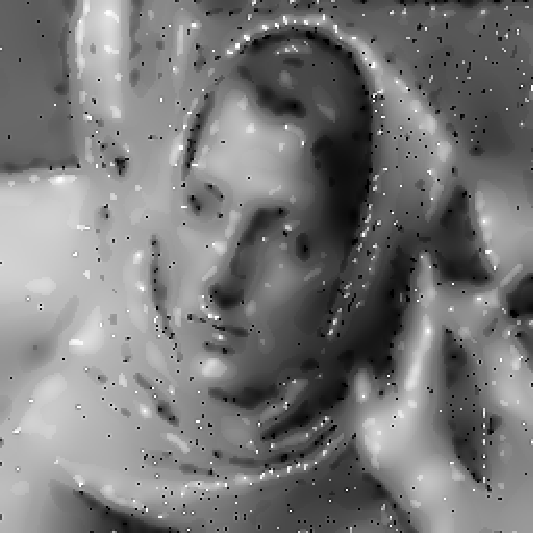}    \\
 (g)  MS-VST + $7/9$& (h) MS-VST + B3 \\
  $NMISE = 0.2391$&$NMISE = 0.3777$\\
  $PSNR = 21.81$&$PSNR = 20.22$
\end{tabular}
}
\caption{{\protect\small Poisson denoising of the Barbara image (image size:
$256 \times 256$).
(a) intensity image (intensity $\in [0.93, 15.73])$;
(b)
Poisson noisy image;
(c) NLMPF ($M=15\times15$, $%
m=21\times21$, $d=0$, $\protect\mu=1$, $NMISE = 0.1150$);
(d) Optimal Weights Filter ($M=15\times15$, $%
m=21\times21$ and $d=0$, $NMISE = 0.1061$);
(e) Poisson NLM ($%
NMISE=0.1207$);
(f) Exact
unbiased inverse + BM3D ($NMISE=0.0863$)
(h) MS-VST + $7/9$ biorthogonal
wavelet ($J = 4$, $FPR = 0.001$, $N_{max}= 5$ iterations, $NMISE = 0.2391$);
(h) MS-VST + B3 isotropic wavelet ($J = 5$, $FPR = 0.001$, $N_{max}= 5$
iterations, $NMISE = 0.3777$).}}
\label{Fig Barbara}
\end{figure}

\begin{figure}[tbp]
\renewcommand{\arraystretch}{0.5} \addtolength{\tabcolsep}{-5pt} \vskip3mm {%
\fontsize{8pt}{\baselineskip}\selectfont
\begin{tabular}{ccc}
\includegraphics[width=0.33\linewidth]{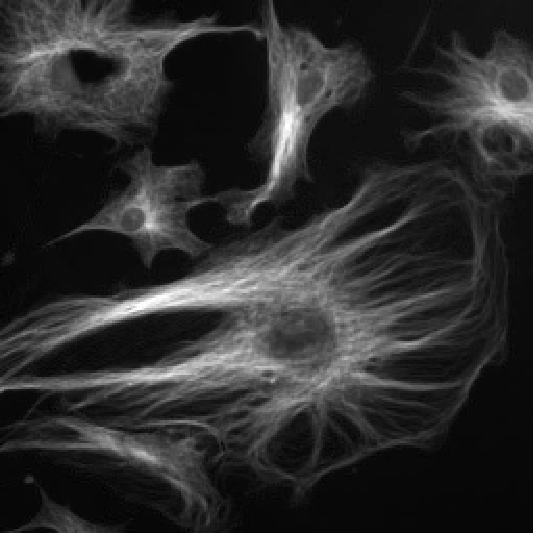} & %
\includegraphics[width=0.33\linewidth]{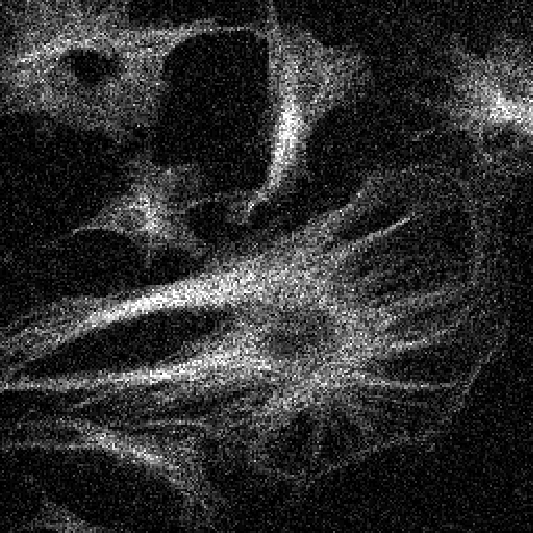} &
\includegraphics[width=0.33\linewidth]{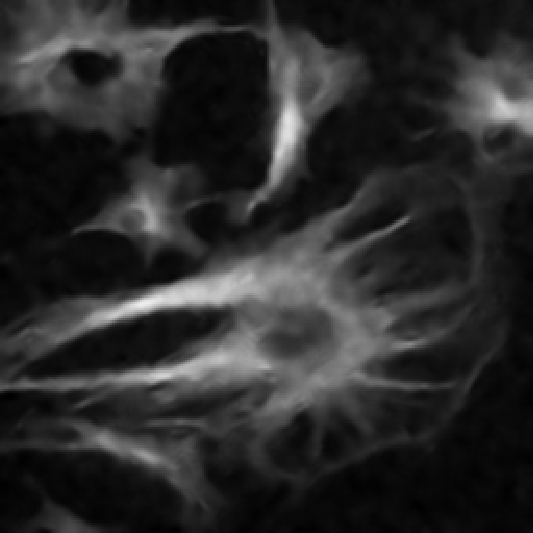}  \\
(a) Original image & (b) Noisy image&
 (c) NLMPF \\&&$NMISE = 0.0785$ \\&&$PSNR = 29.08$ \\
\includegraphics[width=0.33\linewidth]{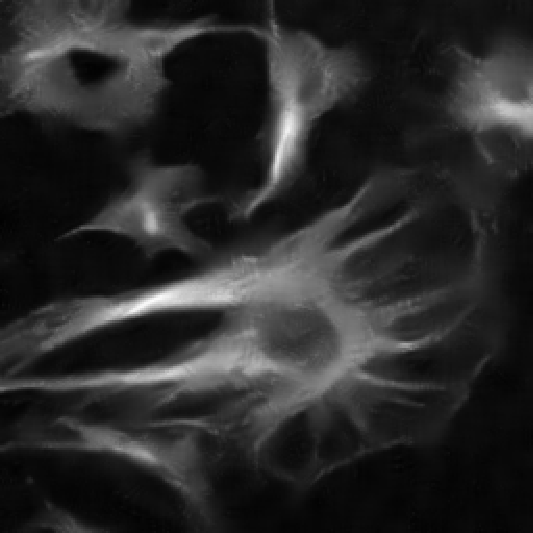} &
\includegraphics[width=0.33\linewidth]{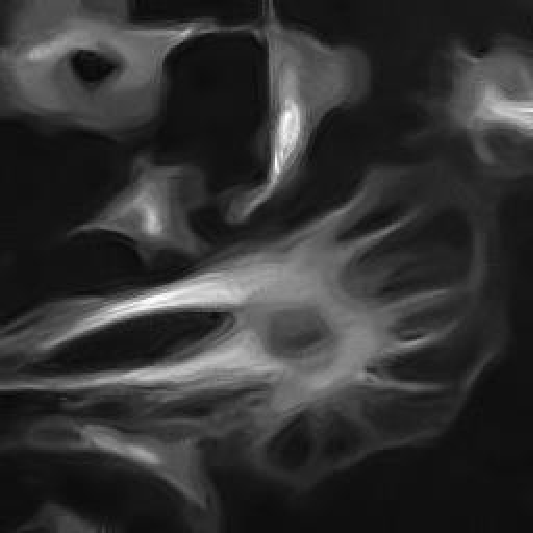} &
\includegraphics[width=0.33\linewidth]{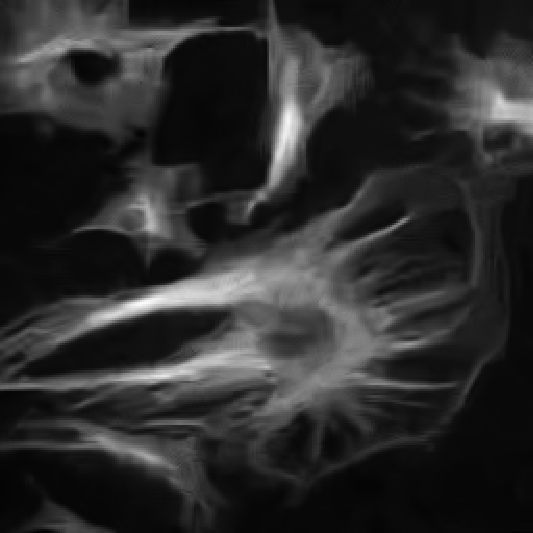} \\
(d)OWPNF& (e) Poisson NLM & (f) EUI+BM3D\\
 $NMISE=0.0794$&$NMISE=0.0816$&$NMISE=0.0643$\\
 $PSNR=29.91$&$PSNR=29.40$&$PSNR=30.18$\\
\includegraphics[width=0.33\linewidth]{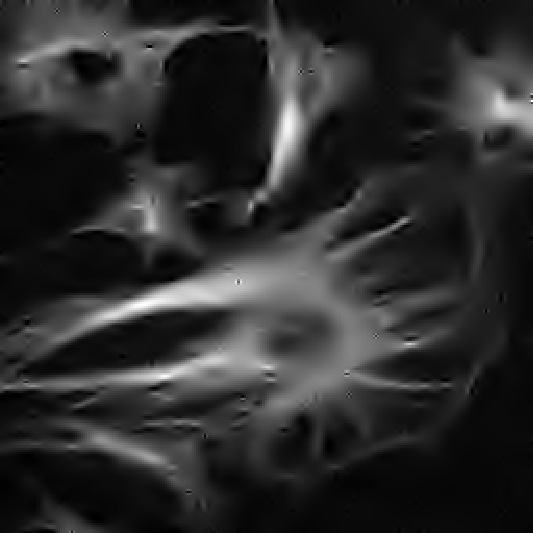} & %
\includegraphics[width=0.33\linewidth]{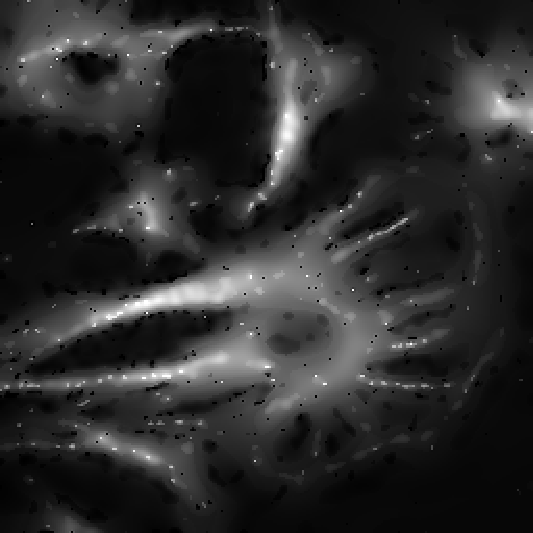}  \\
 (g)  MS-VST + $7/9$& (h) MS-VST + B3 \\
  $NMISE = 0.0909$&$NMISE = 0.1487$ \\
  $PSNR = 28.87$&$PSNR = 26.69$
\end{tabular}
}
\caption{{\protect\small Poisson denoising of fluorescent tubules (image
size: $256 \times 256$).
(a) intensity image (intensity $\in [0.53, 16.93])$;
(b) Poisson noisy image;
(c) NLMPF ($M=7\times7$, $%
m=13\times13$, $d=2$, $\protect\sigma_H=2$, $\protect\mu=1$, $NMISE =
0.0785$);
(d) Optimal Weights Filter ($M=11\times11$, $%
m=17\times17$, $d=1$ and $H=0.6$, $NMISE = 0.0794$);
(e) Poisson NLM ($%
NMISE=0.0816$);
(f) Exact unbiased inverse + BM3D ($NMISE=0.0643$)
(g) MS-VST + $%
7/9$ biorthogonal wavelet ($J = 5$, $FPR = 0.0001$,$N_{max}= 5$ iterations, $%
NMISE = 0.0909$);
(h) MS-VST + B3 isotropic wavelet ($J = 5$, $FPR = 0.001$,
$N_{max}= 10$ iterations, $NMISE = 0.1487$).}}
\label{Fig Cells}
\end{figure}

Table \ref{table1} shows the NMISE values of images reconstructed by NLMPF, OWPNF \citep{jin2012new}, Poisson NLM \citep{deledalle2010poisson}, EUI+BM3D  \cite{makitalo2011optimal}, MS-VST+$7/9$ \cite{ZHANG2008WAVELETS} and  MS-VST+B3 \cite{ZHANG2008WAVELETS}. Our algorithm reach the best in the case of Galaxy$
[0,5]$, while OWPNF reach the best in the case of   Spots$[0.08,4.99]$;
for Ridges$[0.05,0.85]$, Barbara$%
[0.93,15.73]$, and Cells$[0.53,16.93]$, the method EUI+BM3D gives the best
results, but our method is also very competitive.
Table \ref{table2} shows the PSNR values of images reconstructed. Our algorithm also reach the best in the case of Galaxy$
[0,5]$. The method EUI+BM3D have the highest PSNR value. However, the most important evaluation criteria is the visual quality of restored image.
Figures \ref{Fig spots}- \ref{Fig Cells} illustrate the visual quality of
these denoised images. It is obvious that The visual quality of the outputs of our method have high visual quality and many details Remained. For example,
in the case of restored images of Spots (cf. Figures \ref{Fig spots}), our algorithm and OWPNF remain most spots.  We can see clearly $7$ spots at the third column (from left) in  Figures \ref{Fig spots} (c), while EUI+BM3D just remains $4$ spots,  Poisson NLM Makes several spots sticking together, the images restored by MS-VST + 7/9 and MS-VST + B3 are not smooth enough. In the case of Galaxy (cf. Figures  \ref{Fig Galaxy}), visually, our algorithm  best preserves the fine textures.
In the other case, our method also lead to good result visually.

\begin{center}
\begin{table}[tbp]
\caption{A comparison of the denoising performance (NMISE) of several denoising algorithms.}
\begin{center}
\vskip3mm {\footnotesize
\begin{tabular}{|l|r|r|r|r|r|r|}
\hline
Algorithm & Our       & OWPNF       & Poisson & EUI+ & MS-VST   & MS-VST   \\
  &  algorithm &  &  NLM & BM3D &  +$7/9$    &  +B3       \\ \hline
Spots$[0.08,4.99]$ & ${0.0260}$& $\mathbf{0.0259}$&0.0790& $0.0358$ & $0.0602$ & $0.0810$ \\
Galaxy$[0,5]$ & $\mathbf{0.0284}$& ${0.0285}$ & 0.0346 & $0.0297$ & $0.0357$ & $0.0338$      \\
Ridges$[0.05,0.85]$ & $0.0140$& $0.0162$ & 0.0154& $\mathbf{0.0121}$ & $0.0193$ & $0.0416$ \\
Barbara$[0.93,15.73]$ & $0.1150$ & $0.1061$ & 0.1207&$\mathbf{0.0863}$ &$0.2391$ &$0.3777$  \\
Cells$[0.53,16.93]$ & $0.0785$ & $0.0794$ & 0.0816& $\mathbf{0.0643}$ & $0.0909$ & $0.1487$ \\
\hline
\end{tabular}
} \vskip1mm
\end{center}
\label{table1}
\end{table}
\end{center}

\begin{center}
\begin{table}[tbp]
\caption{A comparison of the denoising performance (PSNR, DB) of several denoising algorithms. }
\begin{center}
\vskip3mm {\footnotesize
\begin{tabular}{|l|r|r|r|r|r|r|}
\hline
Algorithm & Our       & OWPNF       & Poisson & EUI+ & MS-VST   & MS-VST   \\
  &  algorithm &  &  NLM & BM3D &  +$7/9$    &  +B3       \\ \hline
Spots$[0.08,4.99]$ & $31.45$& $31.31$&31.12& $\mathbf{31.95}$      & $31.64$ & $30.35$ \\
Galaxy$[0,5]$ & $\mathbf{28.09}$& $27.80$ & 27.77 & $28.04$        & $27.57$ & $27.94$      \\
Ridges$[0.05,0.85]$ & $24.69$& $23.90$ & 24.94& $\mathbf{25.89}$   & $24.49$ & $24.03$ \\
Barbara$[0.93,15.73]$ & $24.71$ & $24.60$ & 24.72&$\mathbf{25.92}$ &$21.81$  &$20.22$  \\
Cells$[0.53,16.93]$ & $29.08$ & $29.91$ & 29.40& $\mathbf{30.18 }$ & $28.87$ & $26.69$ \\
\hline
\end{tabular}
} \vskip1mm
\end{center}
\label{table2}
\end{table}
\end{center}

\section{Conclusion}

In this paper, we have present a new image denoising algorithm to deal with the Poisson noise model, which
is based on the idea of Non-Local Mean. The "Oracle" estimator is obtained
 by   minimized  a very tight upper bound of the Mean Square Error
 with changing the size of search window.
 It help to establish a theorem  to show that  the Non-Local Means Filter can effectively deal with Poisson noise with some modification. As a result,
we successfully construct the new algorithm called Non-Local Means Poisson Filter and demonstrate in statistic theory that the filter converges at the usual optimal rate. The filter is as simple as the classic Non-Local Means and the simulation results show that our filter is very competitive.
 The idea of how to construct an algorithm for Poisson noise model is creative. With our idea, many algorithms  to Remove Gaussian Noise could deal with the Poisson noise with some modification.

\section{\label{sec proofs of mains results} Appendix: Proofs of the main results}

\subsection{\label{proof of th oracle}Proof of Theorem \protect\ref%
{th_oracle}}

Denoting for brevity
\begin{equation}
I_1 = \left(\sum_{x\in \mathbf{U}_{x_0,h}}w^*_h(x)\rho_{f,x_0}(x)\right) ^{2} = \left(\frac{%
\displaystyle \sum_{\|x-x_0\|_{\infty}\leq h} e^{-\frac{\rho_{f,x_0}^2(x)}{H^2(x_0)}
}\rho_{f,x_0}(x)}{\displaystyle \sum_{\|x-x_0\|_{\infty}\leq h}e^{-\frac{\rho_{f,x_0}^2(x)}{%
H^2(x_0)} } }\right)^2,
\label{defi I}
\end{equation}
and
\begin{equation}
I_2
=
 {f}(x_0)\sum_{x\in \mathbf{U}_{x_0,h}}\left(w^*_h(x)\right)^2
 =
 \frac{%
{f}(x_0)\displaystyle \sum_{\|x-x_0\|_{\infty}\leq h} e^{-2\frac{\rho_{f,x_0}^2(x)}{H^2(x_0)}
}}{\left(\displaystyle \sum_{\|x-x_0\|_{\infty}\leq h}e^{-\frac{%
\rho_{f,x_0}^2(x)}{H^2(x_0)} } \right)^2},
\label{defi I2}
\end{equation}
then we have
\begin{equation}
g(w^*_h)=I_1+I_2.  \label{MSE bound}
\end{equation}
The conditions (\ref{condition of Hx}) and $\gamma > \sqrt{2}Lh^{\beta}$ imply that for $x\in U_{x_0,h}$, we have
\begin{equation}
\frac{L^2\|x-x_0\|_{\infty}^{2\beta}}{H^2(x)}\leq \frac{L^2h^{2\beta}}{\gamma^2}\leq  \frac{1}{2}.
\label{ineq frac LhH}
\end{equation}
Noting that  $e^{-\frac{t^2}{H^2(x_0)}}$, $t\in [0,\gamma/\sqrt 2)$
is decreasing, and using one term Taylor expansion, the inequality (\ref{ineq frac LhH}) implies that
\begin{eqnarray}
\displaystyle \sum_{\|x-x_0\|_{\infty}\leq h}e^{-\frac{%
\rho_{f,x_0}^2(x)}{H^2(x_0)} }
&\geq&
\sum_{\|x-x_0\|_{\infty}\leq h}e^{-\frac{L^2\|x-x_0\|_{\infty}^{2\beta}}{%
H^2(x_0)}}
\geq
\sum_{\|x-x_0\|_{\infty}\leq h} \left(1-\frac{%
L^2\|x-x_0\|_{\infty}^{2\beta}}{H^2(x_0)}\right)
\nonumber \\&\geq&
2h^2n.
\label{weight taylor}
\end{eqnarray}
Considering that $te^{-\frac{t^2}{H^2(x_0)}}$, $t\in [0, \gamma/\sqrt 2)$ is
increasing function,
\begin{eqnarray}
\sum_{\|x-x_0\|_{\infty}\leq h} e^{-\frac{\rho_{f,x_0}^2(x)}{H^2(x_0)}
}\rho_{f,x_0}(x)
&\leq&
\sum_{\|x-x_0\|_{\infty}\leq h} L\|x-x_0\|^{\beta}_{\infty}e^{-\frac{%
L^2\|x-x_0\|_{\infty}^{2\beta}}{H^2(x_0)}}
\nonumber\\&\leq&
\sum_{\|x-x_0\|_{\infty}\leq
h} L\|x-x_0\|^{\beta}_{\infty}\leq 4Lh^{\beta+2}n.
\label{Inequ E}
\end{eqnarray}
The above three inequalities (\ref{defi I}), (\ref{weight taylor}) and (\ref{Inequ E}) imply that
\begin{equation}
I_1 \leq 4L^2h^{2\beta}.  \label{I1 bound}
\end{equation}
Taking into account the inequality
$$
\sum_{\|x-x_0\|_{\infty}\leq h} e^{-2\frac{\rho_{f,x_0}^2(x)}{H^2(x_0)}
} \leq\sum_{\|x-x_0\|_{\infty}\leq h}1=4h^2n,
$$
(\ref{defi I2}) and (\ref{weight taylor}), it is easily seen that
\begin{equation}
I_2\leq \frac{\Gamma}{h^2n}.  \label{I2 bound}
\end{equation}
Combining (\ref{MSE bound}), (\ref{I1 bound}), and (\ref{I2 bound}), we give
\begin{equation}
g(w_h^*)\leq 4L^2h^{2\beta}+\frac{\Gamma}{h^2n}.  \label{upper bound ineq}
\end{equation}
Let $h$ minimize the latter term of the above inequality (\ref{upper bound ineq}). Then
$$
8\beta L ^2 h^{2\beta-1}-\frac{2\Gamma}{h^3n}=0
$$
from which we infer that
\begin{equation}
h=\left(\frac{\Gamma}{4\beta L^2}\right)^{\frac{1}{2\beta+2}}n^{-\frac{1}{%
2\beta+2}}.  \label{h value}
\end{equation}
Substituting (\ref{h value}) to (\ref{upper bound ineq}) leads to
$$
g(w_h^*)\leq \frac{2^{\frac{2\beta + 6}{2\beta +2}} \Gamma^{\frac{2\beta}{%
2\beta +2}} L^{\frac{4}{2\beta +2}} }{\beta^{\frac{2\beta}{2\beta +2}}} n^{-%
\frac{2\beta}{2\beta+2}}.
$$
Therefore (\ref{upper_bound}) implies (\ref{rate_oracle}).

\subsection{\label{proof of similar condition} Proof of Theorem \protect\ref%
{th similar function}}
We shall use following lemma to finish the Proof of Theorem \protect\ref{th similar function}.
The  lemma can be deduced form the results in Borovkov %
\cite{borovkov2000estimates}, see also Merlevede, Peligrad and Rio \cite{merleve2010bernstein}

\begin{lemma}
\label{Lemma Borovk} If, for some $\delta >0,\gamma \in (0,1)$ and $K>1$ we
have
$$
\sup \mathbb{E}\exp \left(\delta \left\vert X_{i}\right\vert ^{\gamma }\right) \leq
K,\;i=1,...,n,
$$%
then there are two positive constants $c_{1}$ and $c_{2}$ depending only on $%
\delta ,$ $\gamma $ and $K$ such that, for any $t>0,$
$$
\mathbb{P}\left(\sum_{i=1}^{n}X_{i}\geq t\right) \leq \exp \left(-c_{1}t^{2}/n\right) +n\exp \left(-c_{2}t^{\gamma }\right) .
$$
\end{lemma}

{\bf Proof of Theorem \protect\ref%
{th similar function}. } Recall that $M$ and $m$ is given by (\ref{defi M}) and (\ref{defi m}) respectively. Consider that
\begin{equation}
|\widehat{\rho }_{x_0}^2(x)-\rho_{f,x_0}^2(x)|\leq \left\vert \frac{1}{m}S(x)+\frac{1}{m}%
R(x)\right\vert ,  \label{express rho}
\end{equation}%
where
$$
S(x)=\sum_{y\in \mathbf{U}_{x_{0},\eta }} Z(y),
$$%
\begin{equation}
Z(y)=(Y(y)-Y(T_x y))^{2}-(f(y)-f(T_x y))^{2}-f(y)-f(T_x y),
\end{equation}%
and
$$
R(x)=\sum_{y\in \mathbf{U}_{x_{0},\eta }}\left((f(y)-f(T_x y))^{2}+f(y)+f(T_x y)\right) -(f(x_{0})-f(x))^{2}-2\overline{f}(x_{0}).
$$

Since $Y(x)$ has the Poisson distribution, with mean $f(x)$ and variance $%
f(x)$,
\begin{equation}
\mathbb{E}\left(e^{Y(x)}\right)=\sum_{k=0}^{+\infty }e^{k}\frac{f^{k}(x)e^{-f(x)}}{k!}%
=ef(x)e^{(e-1)f(x)}\leq e\Gamma e^{(e-1)\Gamma }.  \label{grand number}
\end{equation}%
From the inequality (\ref{grand number}),  we easily deduce
\begin{equation}
\sup \mathbb{E}\left(e^{|Z(y)|^{1/2}}\right) \leq \sup \mathbb{E}\left(
e^{Y(y)+Y(T_x y)+2\Gamma +2\sqrt{\Gamma }}\right) \leq (e\Gamma
)^{2}e^{2e\Gamma +2\sqrt{\Gamma }}  \label{upper bound}
\end{equation}%
By Lemma \ref{Lemma Borovk},
we see that there are two positive constants $c_{5}$ and $c_{6}$ such that for any $z>0$,
\begin{equation}
\mathbb{P}\left(\frac{1}{m}|S(x)|\geq \frac{z}{\sqrt{m}}\right) \leq \exp
(-c_{5}z^{2})+m\exp (-c_{6}(\sqrt{m}z)^{\frac{1}{2}}).
\label{inequality important}
\end{equation}%
Considering $m=(2N\eta+1)^2$ and $\eta=c_1 n^{-\alpha}(\frac{(1-\beta)^+)}{2\beta+2}<\alpha< \frac{1}{2}$, we have $m=c'_1 n ^{1-2\alpha}(1+o(1))$. Therefore, substituting $z=\sqrt{\frac{1}{c_{5}}\ln n^{2}}$ into the inequality (\ref%
{inequality important}), we see that for $n$ large enough,
$$
\mathbb{P}\left(\frac{1}{m}\left\vert S(x)\right\vert \geq \frac{\sqrt{%
\frac{1}{c_{5}}\ln n^{2}}}{\sqrt{m}}\right) \leq 2\exp \left(-\ln
n^{2}\right) =\frac{2}{n^{2}}.
$$%
From this inequality we easily deduce that
$$
\mathbb{P}\left(\max_{x\in \mathbf{U}_{x_{0},h}}\frac{1}{m}
\left\vert S(x)\right\vert \geq \frac{\sqrt{\frac{1}{c_{5}}\ln n^{2}}}{\sqrt{%
m}}\right) \leq \sum_{x\in \mathbf{U}_{x_{0},h}}\mathbb{P}\left(\frac{1}{m}\left\vert S(x)\right\vert \geq \frac{\sqrt{\frac{1}{c_{5}}\ln
n^{2}}}{\sqrt{m}}\right) \leq \frac{2}{n}.
$$%
We arrive at
\begin{equation}
\mathbb{P}\left(\mathbf{B}\right) \leq c_{7}n^{-1},  \label{ineq p b}
\end{equation}%
where $\mathbf{B}=\{\max_{x\in \mathbf{U}_{x_{0},h}}\frac{1}{m}
\left\vert S(x)\right\vert <c_{8}n^{\alpha -\frac{1}{2}}\sqrt{\ln n}\}$ and $%
c_{8}$ is a constant depending only on $\beta $ and $L$. It is easy to see
that
\begin{equation}
R(x)=O\left(n^{\alpha -\frac{1}{2}}\right) .  \label{upper bound Rx}
\end{equation}%
In the set $\mathbf{B}$, the inequality (\ref{upper bound Rx}) implies that
\begin{equation}
\max_{x\in \mathbf{U}_{x_{0},h}}|\widehat{\rho }_{x_0}^{2}(x)-\rho_{f,x_0}^2(x)|\leq
c_{8}n^{-\frac{\beta }{2\beta +2}}\sqrt{\ln n}+O\left(n^{\alpha -\frac{1}{2}%
}\right) =O(n^{-\frac{\beta }{2\beta +2}}\sqrt{\ln n}).  \label{ineq rhorho}
\end{equation}%
Combining (\ref{ineq p b}) and (\ref{ineq rhorho}), we obtain (\ref{rate
similar function}).

\subsection{\label{proof of rate estimator}Proof of Theorem \protect\ref{th
rate estimator}}

Taking into account (\ref{estimator similar}), (\ref{estimate002}), and the
independence of $\epsilon (x)$, we have
\begin{equation}
\mathbb{E}\{|\widehat{f}'_{h }(x_{0})-f(x_{0})|^{2}\big|Y(x),x\in
\mathbf{I}''_{x_{0}}\}\leq g^{\prime }(\widehat{w}_h),  \label{MSE estimator}
\end{equation}%
where
$$
g^{\prime }(w)=\left(\sum_{x\in \mathbf{U}'_{x_{0},h}}{w}%
(x)\rho_{f,x_0}(x)\right) ^{2}+\overline{f}'(x_{0})\sum_{x\in \mathbf{I}'_{x_{0}}}{w}^{2}(x).
$$%
By the proof of Theorem \ref{th_oracle}, we obtain
\begin{equation}
g^{\prime }(w^{\ast }_{h})\leq \frac{3}{2}\left(\frac{2^{\frac{2\beta +6}{%
2\beta +2}}\Gamma ^{\frac{2\beta }{2\beta +2}}L^{\frac{4}{2\beta +2}}}{\beta
^{\frac{2\beta }{2\beta +2}}}n^{-\frac{2\beta }{2\beta +2}}\right) .
\label{gx bound}
\end{equation}%
By Theorem \ref{th similar function} and its proof, for $\widehat{\rho}'_{x_0}$
is defined by (\ref{estimator similar}), there is a constant $c_{2}$
such that
\begin{equation}
\mathbb{P}\left\{ \max_{x\in \mathbf{U}'_{x_{0},h}}\left\vert \widehat{\rho }_{x_0}
^{'2}(x)-\rho_{f,x_0} ^{2}(x)\right\vert \geq c_{2}n^{\alpha -\frac{1}{2}}\sqrt{\ln n}%
\right\} =O\left( n^{-1}\right) .  \label{rate01 similar function}
\end{equation}
Let $\mathbf{B}=\left\{ \max_{x\in \mathbf{U}'_{x_{0},h}}\left\vert
\widehat{\rho }_{x_0}^{'2}(x)-\rho_{f,x_0}^2(x)\right\vert \leq c_{2}n^{\alpha -\frac{1}{%
2}}\ln n\right\} $. On the set $\mathbf{B}$, we have $\rho_{f,x_0}^2(x)-c_{2}n^{\alpha -\frac{1}{2}}\sqrt{\ln n}<\widehat{\rho }_{x_0}^{'2}(x)<\rho_{f,x_0}^2(x)+c_{2}n^{\alpha -\frac{1}{2}}\sqrt{\ln n}$, from which we infer that
\begin{eqnarray*}
\widehat{w}(x)& =&\frac{e^{-\frac{\widehat{\rho }_{x_0}^{ '2}(x)}{H^{2}(x_{0})}}}{%
\sum_{x^{\prime }\in \mathbf{U}_{x_{0},h}^{\prime }}e^{-\frac{\widehat{\rho }_{x_0}%
^{'2}(x^{\prime })}{H^{2}(x_{0})}}}\leq \frac{e^{-\frac{{\rho }_{f,x_0}^{2}(x)-c_{2}n^{\alpha -\frac{1}{2}}\sqrt{\ln n}}{H^{2}(x_{0})}}}{%
\sum_{x^{\prime }\in \mathbf{U}_{x_{0},h}^{\prime }}e^{-\frac{{\rho }_{f,x_0}^{2}(x^{\prime })+c_{2}n^{\alpha -\frac{1}{2}}\sqrt{\ln n}}{H^{2}(x_{0})}}}
\\
& \leq &
\frac{e^{-\frac{{\rho }_{f,x_0}^{2}(x)}{H^{2}(x_{0})}}\left(1+2\frac{%
c_{2}n^{\alpha -\frac{1}{2}}\sqrt{\ln n}}{H^{2}(x_{0})}\right) }{%
\sum_{x^{\prime }\in \mathbf{U}_{x_{0},h}^{\prime }}e^{-\frac{{\rho }_{f,x_0}^{2}(x^{\prime })}{H^{2}(x_{0})}}\left(1-\frac{c_{2}n^{\alpha -\frac{1}{2}}%
\sqrt{\ln n}}{H^{2}(x_{0})}\right) }
\\
& =&
\left(\frac{1+\frac{2c_{2}n^{\alpha -\frac{1}{2}}\sqrt{\ln n}}{%
H^{2}(x_{0})}}{1-\frac{c_{2}n^{\alpha -\frac{1}{2}}\sqrt{\ln n}}{H^{2}(x_{0})%
}}\right) \frac{e^{-\frac{{\rho }_{f,x_0}^{2}(x)}{H^{2}(x_{0})}}}{\sum_{x^{\prime
}\in \mathbf{U}_{x_{0},h}^{\prime }}e^{-\frac{{\rho }_{f,x_0}^{2}(x^{\prime })}{%
H^{2}(x_{0})}}}
\\
& =&
\left(\frac{1+\frac{2c_{2}n^{\alpha -\frac{1}{2}}\sqrt{\ln n}}{%
H^{2}(x_{0})}}{1-\frac{c_{2}n^{\alpha -\frac{1}{2}}\sqrt{\ln n}}{H^{2}(x_{0})%
}}\right) w^*_h(x).
\end{eqnarray*}
This implies that
$$
g^{\prime }(\widehat{w}_{h})\leq \left(\frac{1+\frac{2c_{2}n^{\alpha -\frac{%
1}{2}}\sqrt{\ln n}}{H^{2}(x_{0})}}{1-\frac{c_{2}n^{\alpha -\frac{1}{2}}\sqrt{%
\ln n}}{H^{2}(x_{0})}}\right) ^{2}g'(w^*_h).
$$%
The condition $4c_{2}n^{\alpha -\frac{1}{2}}\sqrt{\ln n}<\gamma\leq H^{2}(x_{0})$
implies that
$$
\left(\frac{1+\frac{2c_{2}n^{\alpha -\frac{1}{2}}\sqrt{\ln n}}{H^{2}(x_{0})}%
}{1-\frac{c_{2}n^{\alpha -\frac{1}{2}}\sqrt{\ln n}}{H^{2}(x_{0})}}\right)
^{2}\leq 2.
$$%
Consequently, (\ref{MSE estimator}) becomes
\begin{equation}
\mathbb{E}\left(|\widehat{f}'_{h }(x_{0})-f(x_{0})|^{2}\big|Y(x),x\in
\mathbf{I}''_{x_{0}},\mathbf{B}\right)\leq 2g^{\prime  }(w^*_h).  \label{expection bound}
\end{equation}%
Since the function $f$ satisfies the H\"{o}lder condition,
\begin{equation}
\mathbb{E}\left(|\widehat{f}'_{h}(x_{0})-f(x_{0})|^{2}\big|Y(x),x\in \mathbf{%
I}''_{x_{0}}\right) <g'(\widehat{w}_{h})\leq c_{9},
\label{expection bound K}
\end{equation}%
for a constant $c_{9}>0$ depending only on $\beta $ and $L$.
Combining (\ref{rate similar function}), (\ref{expection bound}), and (\ref%
{expection bound K}), we have
\begin{eqnarray*}
\mathbb{E}&& \left(|\widehat{f}'_{h}(x_{0})-f(x_{0})|^{2}\big|Y(x),x\in
\mathbf{I}''_{x_{0}}\right)
\\
&=& \mathbb{E}\left(|\widehat{f}'_{h }(x_{0})-f(x_{0})|^{2}\big|Y(x),x\in
\mathbf{I}''_{x_{0}},\mathbf{B}\right)\mathbb{P}(\mathbf{B})
\\
&& +\mathbb{E}\left(|\widehat{f}'_{h }(x_{0})-f(x_{0})|^{2}\big|Y(x),x\in
\mathbf{I}''_{x_{0}},\overline{\mathbf{B}}\right)\mathbb{P}(\overline{\mathbf{B}})
\\
&\leq & 2g^{\prime  }(w^*_h)+O\left(n^{-1}\right)
\end{eqnarray*}

%
Now, the assertion of the theorem is obtained easily if we take into account
(\ref{gx bound}).



\begin{thebibliography}{10}

\bibitem{Abraham2007significant}
I.~Abraham, R.~Abraham, A.~Desolneux, and S.~Li-Thiao-Te.
\newblock Significant edges in the case of non-stationary gaussian noise.
\newblock {\em Pattern recognition}, 40(11):3277--3291, 2007.

\bibitem{aharon2006rm}
M.~Aharon, M.~Elad, and A.~Bruckstein.
\newblock $ rm k $-svd: An algorithm for designing overcomplete dictionaries
  for sparse representation.
\newblock {\em IEEE Trans. Signal Process.}, 54(11):4311--4322, 2006.

\bibitem{alter2006intensity}
F.~Alter, Y.~Matsushita, and X.~Tang.
\newblock An intensity similarity measure in low-light conditions.
\newblock {\em Computer Vision--ECCV 2006}, pages 267--280, 2006.

\bibitem{ANSCOMBE1948TRANSFORMATION}
F.J. Anscombe.
\newblock The transformation of poisson, binomial and negative-binomial data.
\newblock {\em Biometrika}, 35(3/4):246--254, 1948.

\bibitem{beck2009fast}
A.~Beck and M.~Teboulle.
\newblock Fast gradient-based algorithms for constrained total variation image
  denoising and deblurring problems.
\newblock {\em IEEE Trans. Image Process.}, 18(11):2419--2434, 2009.

\bibitem{borovkov2000estimates}
A.A. Borovkov.
\newblock Estimates for the distribution of sums and maxima of sums of random
  variables without the cramer condition.
\newblock {\em Siberian Mathematical Journal}, 41(5):811--848, 2000.

\bibitem{boulanger2008non}
J.~Boulanger, J.B. Sibarita, C.~Kervrann, and P.~Bouthemy.
\newblock Non-parametric regression for patch-based fluorescence microscopy
  image sequence denoising.
\newblock In {\em in Proc. of IEEE Int. Symp. on Biomedical Imaging: From Nano
  to Macro, ISBI¡¯2008}, pages 748--751. IEEE, 2008.

\bibitem{buades2005review}
A.~Buades, B.~Coll, and J.M. Morel.
\newblock A review of image denoising algorithms, with a new one.
\newblock {\em SIAM Journal on Multiscale Modeling and Simulation},
  4(2):490--530, 2005.

\bibitem{buades2009note}
T.~Buades, Y.~Lou, JM~Morel, and Z.~Tang.
\newblock {A note on multi-image denoising}.
\newblock In {\em Int. workshop on Local and Non-Local Approximation in Image
  Processing}, pages 1--15, August 2009.

\bibitem{deledalle2010poisson}
C.A. Deledalle, F.~Tupin, and L.~Denis.
\newblock Poisson nl means: Unsupervised non local means for poisson noise.
\newblock In {\em IEEE Int. Conf. on Image Process. (ICIP), 2010 17th}, pages
  801--804. IEEE, 2010.

\bibitem{donoho1994ideal}
D.L. Donoho and J.M. Johnstone.
\newblock {Ideal spatial adaptation by wavelet shrinkage}.
\newblock {\em Biometrika}, 81(3):425, 1994.

\bibitem{fan1993local}
J.~Fan.
\newblock Local linear regression smoothers and their minimax efficiencies.
\newblock {\em The Annals of Statistics}, pages 196--216, 1993.

\bibitem{FanGijbels1996}
J.Q. Fan and I.~Gijbels.
\newblock {Local polynomial modelling and its applications}.
\newblock In {\em Chapman \& Hall, London}, 1996.

\bibitem{FRYZLEWICZ2007GOES}
P.~Fryzlewicz, V.~Delouille, and G.P. Nason.
\newblock Goes-8 x-ray sensor variance stabilization using the multiscale
  data-driven haar--fisz transform.
\newblock {\em J. Roy. Statist. Soc. ser. C}, 56(1):99--116, 2007.

\bibitem{FRYZLEWICZ2004HAAR}
P.~Fryzlewicz and G.P. Nason.
\newblock A haar-fisz algorithm for poisson intensity estimation.
\newblock {\em J. Comp. Graph. Stat.}, 13(3):621--638, 2004.

\bibitem{hammond2008image}
D.K. Hammond and E.P. Simoncelli.
\newblock Image modeling and denoising with orientation-adapted gaussian scale
  mixtures.
\newblock {\em IEEE Trans. Image Process.}, 17(11):2089--2101, 2008.

\bibitem{hirakawa2006image}
K.~Hirakawa and T.W. Parks.
\newblock Image denoising using total least squares.
\newblock {\em IEEE Trans. Image Process.}, 15(9):2730--2742, 2006.

\bibitem{jansen2006multiscale}
M.~Jansen.
\newblock Multiscale poisson data smoothing.
\newblock {\em J. Roy. Statist. Soc. B}, 68(1):27--48, 2006.

\bibitem{jin2012new}
Qiyu Jin, Ion Grama, and Quansheng Liu.
\newblock A new poisson noise filter based on weights optimization.
\newblock {\em arXiv preprint arXiv:1201.5968}, 2012.

\bibitem{JinGramaLiuowf}
Q.Y. Jin, I.~Grama, and Q.S. Liu.
\newblock Removing gaussian noise by optimization of weights in non-local
  means.
\newblock {\em http://arxiv.org/abs/1109.5640}.

\bibitem{Katkovnik2010local}
V.~Katkovnik, A.~Foi, K.~Egiazarian, and J.~Astola.
\newblock {From local kernel to nonlocal multiple-model image denoising}.
\newblock {\em Int. J. Comput. Vis.}, 86(1):1--32, 2010.

\bibitem{kervrann2006optimal}
C.~Kervrann and J.~Boulanger.
\newblock {Optimal spatial adaptation for patch-based image denoising}.
\newblock {\em IEEE Trans. Image Process.}, 15(10):2866--2878, 2006.

\bibitem{kolaczyk2005multiscale}
E.D. Kolaczyk and R.D. Nowak.
\newblock Multiscale generalised linear models for nonparametric function
  estimation.
\newblock {\em Biometrika}, 92(1):119, 2005.

\bibitem{le2007variational}
T.~Le, R.~Chartrand, and T.~J. Asaki.
\newblock A variational approach to reconstructing images corrupted by poisson
  noise.
\newblock {\em Journal of Mathematical Imaging and Vision}, 27(3):257--263,
  2007.

\bibitem{lefkimmiatis2009bayesian}
S.~Lefkimmiatis, P.~Maragos, and G.~Papandreou.
\newblock Bayesian inference on multiscale models for poisson intensity
  estimation: Applications to photon-limited image denoising.
\newblock {\em IEEE Trans. Image Process.}, 18(8):1724--1741, 2009.

\bibitem{luisier2010fast}
F.~Luisier, C.~Vonesch, T.~Blu, and M.~Unser.
\newblock Fast interscale wavelet denoising of poisson-corrupted images.
\newblock {\em Signal Process.}, 90(2):415--427, 2010.

\bibitem{mairal2008learning}
J.~Mairal, G.~Sapiro, and M.~Elad.
\newblock Learning multiscale sparse representations for image and video
  restoration.
\newblock {\em SIAM Multiscale Modeling and Simulation}, 7(1):214--241, 2008.

\bibitem{MAKITALO2009INVERSION}
M.~Makitalo and A.~Foi.
\newblock On the inversion of the anscombe transformation in low-count poisson
  image denoising.
\newblock In {\em Proc. Int. Workshop on Local and Non-Local Approx. in Image
  Process., LNLA 2009, Tuusula, Finland}, pages 26--32. IEEE, 2009.

\bibitem{makitalo2011optimal}
M.~Makitalo and A.~Foi.
\newblock Optimal inversion of the anscombe transformation in low-count poisson
  image denoising.
\newblock {\em IEEE Trans. Image Process.}, 20(1):99--109, 2011.

\bibitem{mandel1982use}
J.~Mandel.
\newblock Use of the singular value decomposition in regression analysis.
\newblock {\em The American Statistician}, 36(1):15--24, 1982.

\bibitem{merleve2010bernstein}
F.~Merlev{\`e}de, M.~Peligrad, and E.~Rio.
\newblock A bernstein type inequality and moderate deviations for weakly
  dependent sequences.
\newblock {\em Probab. Theory Related Fields}, 2010.

\bibitem{moon2009three}
I.~Moon and B.~Javidi.
\newblock Three dimensional imaging and recognition using truncated photon
  counting model and parametric maximum likelihood estimator.
\newblock {\em Optics express}, 17(18):15709--15715, 2009.

\bibitem{nowak2000statistical}
R.D. Nowak and E.D. Kolaczyk.
\newblock A statistical multiscale framework for poisson inverse problems.
\newblock {\em IEEE Trans. Info. Theory}, 46(5):1811--1825, 2000.

\bibitem{polzehl2006propagation}
J.~Polzehl and V.~Spokoiny.
\newblock {Propagation-separation approach for local likelihood estimation}.
\newblock {\em Probab. Theory Rel.}, 135(3):335--362, 2006.

\bibitem{portilla2003image}
J.~Portilla, V.~Strela, M.J. Wainwright, and E.P. Simoncelli.
\newblock Image denoising using scale mixtures of gaussians in the wavelet
  domain.
\newblock {\em IEEE Trans. Image Process.}, 12(11):1338--1351, 2003.

\bibitem{roth2009fields}
S.~Roth and M.J. Black.
\newblock Fields of experts.
\newblock {\em Int. J. Comput. Vision}, 82(2):205--229, 2009.

\bibitem{terrell1992variable}
G.~R. Terrell and D.~W. Scott.
\newblock Variable kernel density estimation.
\newblock {\em The Annals of Statistics}, pages 1236--1265, 1992.

\bibitem{van1996comparing}
G.M.P. van Kempen, H.T.M. van~der Voort, J.G.J. Bauman, and K.C. Strasters.
\newblock Comparing maximum likelihood estimation and constrained
  tikhonov-miller restoration.
\newblock {\em IEEE Engineering in Medicine and Biology Magazine},
  15(1):76--83, 1996.

\bibitem{ZHANG2008WAVELETS}
B.~Zhang, J.M. Fadili, and J.L. Starck.
\newblock Wavelets, ridgelets, and curvelets for poisson noise removal.
\newblock {\em IEEE Trans. Image Process.}, 17(7):1093--1108, 2008.

\end{thebibliography}

\end{document}